\documentclass[twocolumn]{aastex631}
\usepackage{xcolor}
\usepackage{subfloat}
\usepackage{multirow}
\usepackage{soul}
\usepackage{graphicx}	% Including figure files
\usepackage{amsmath}	% Advanced maths commands

\usepackage{amssymb}	% Extra maths symbols
\usepackage{cleveref}   % cross-cref
\usepackage{ulem}
\usepackage{multirow}
\usepackage{longtable}
\usepackage{comment}

\begin{document}

\title{Investigation of Transit Timing and an Optical Transmission Spectrum of the Hot Jupiter WASP-11~b}

\author[0000-0001-7234-7167]{Napaporn A-thano}
\affiliation{National Astronomical Research Institute of Thailand, 260 Moo 4, Donkaew, Mae Rim, Chiang Mai, 50180, Thailand}
\email{napaporn@narit.or.th}

\author[0000-0003-3251-3583]{Supachai Awiphan}
\affiliation{National Astronomical Research Institute of Thailand, 260 Moo 4, Donkaew, Mae Rim, Chiang Mai, 50180, Thailand}
\email{supachai@narit.or.th}

\author[0000-0002-1743-4468]{Eamonn Kerins}
\affiliation{Jodrell Bank Centre for Astrophysics, University of Manchester, Oxford Road, Manchester, M13 9PL, UK}

\author[0000-0003-1143-0877]{Akshay Priyadarshi}
\affiliation{Jodrell Bank Centre for Astrophysics, University of Manchester, Oxford Road, Manchester, M13 9PL, UK}

\author[0000-0003-0356-0655]{Iain McDonald}
\affiliation{Jodrell Bank Centre for Astrophysics, University of Manchester, Oxford Road, Manchester, M13 9PL, UK}
\affiliation{Department of Physical Sciences, The Open University, Walton Hall, Milton Keynes, MK7 6AA, UK}

\author{Orarik Tasuya}
\affiliation{National Astronomical Research Institute of Thailand, 260 Moo 4, Donkaew, Mae Rim, Chiang Mai, 50180, Thailand}

\author{Ronnakrit Rattanamala}
\affiliation{Department of Physics and General Science, Faculty of Science and Technology, \\ Nakhon Ratchasima Rajabhat University, Nakhon Ratchasima, 30000, Thailand}

\author[0000-0001-7359-3300]{Ing-Guey Jiang}
\affiliation{Department of Physics and Institute of Astronomy, National Tsing-Hua University, Hsinchu 30013, Taiwan}

\author[0000-0001-8657-1573]{Yogesh C. Joshi}
\affiliation{Aryabhatta Research Institute of Observational Sciences (ARIES), Manora Peak, Nainital 263001, India}

\author[0000-0002-6039-8212]{Fan Yang}
\affil{D\'epartement d'Astrophysique/AIM, CEA/IRFU, CNRS/INSU, Univ. Paris-Saclay, Univ. de Paris, 91191 Gif-sur-Yvette, France\\}

\author{Ida Janiak}
\affiliation{Jodrell Bank Centre for Astrophysics, University of Manchester, Oxford Road, Manchester, M13 9PL, UK}

\author{Patcharawee Munsaket}
\affiliation{School of Physics, Institute of Science, Suranaree University of Technology, 111 University Ave., \\ Suranaree, Nakhon Ratchasima 30000, Thailand}

\author[0000-0001-5162-4225]{Yasir Abdul Qadir}
\affil{Department of Physics and Astronomy, FI-20014 University of Turku, Finland\\}

\author{Smanchan Chandaiam}
\affiliation{National Astronomical Research Institute of Thailand, 260 Moo 4, Donkaew, Mae Rim, Chiang Mai, 50180, Thailand}

\author{Boonyarit Choonhakit}
\affiliation{National Astronomical Research Institute of Thailand, 260 Moo 4, Donkaew, Mae Rim, Chiang Mai, 50180, Thailand}

\author{Suwanit Wutsang}
\affiliation{National Astronomical Research Institute of Thailand, 260 Moo 4, Donkaew, Mae Rim, Chiang Mai, 50180, Thailand}

\author{Boonrucksar Soonthornthum}
\affiliation{National Astronomical Research Institute of Thailand, 260 Moo 4, Donkaew, Mae Rim, Chiang Mai, 50180, Thailand}

\author{Vik S Dhillon}
\affiliation{Department of Physics and Astronomy, University of Sheffield, Sheffield, S3 7RH, UK}
\affiliation{Instituto de Astrofísica de Canarias, E-38205 La Laguna, Tenerife, Spain}



\begin{abstract}

WASP-11~b/HAT-P-10~b is an inflated hot Jupiter, which has a low density that makes it a good target for atmospheric studies using the transmission spectroscopy technique. In this work, we present 31 new transit light curves of WASP-11~b/HAT-P-10~b, obtained through the SPEARNET network. These data were analyzed along with previously published ground-based observations and space-based data from \texttt{TESS}. We refine the planetary parameters of WASP-11~b/HAT-P-10~b and perform a transit timing analysis using data spanning 16 years. The updated ($O-C$) diagram shows no significant evidence of orbital decay. The TTV analysis reveals no significant signals indicative of additional planets. Atmospheric analysis using multi-band optical observations indicates a strong Rayleigh scattering slope in the transmission spectra, which may originate from the planetary atmosphere itself or be influenced by contamination such as stellar activity or light from the companion star.

\end{abstract}



\section{Introduction} 
\label{sec:intro}

In 2009, the transiting exoplanet WASP-11~b/HAT-P-10~b was independently discovered by two ground-based surveys: the Wide-Angle Search for Planets (WASP; \citet{west2009}) and the Hungarian-made Automated Telescope Network (HATNet; \citet{bakos2009}). The planet orbits an early K-dwarf star, WASP-11A/HAT-P-10A ($V$ = 12, distance = 129.88 $\pm$ 0.98 pc; based on a Gaia DR3 parallax of 7.6997 $\pm$ 0.0579 mas), with a period of 3.722 days. \citet{bakos2009} reported the discovery using the HAT-10 telescope as part of the HATNet survey and derived a planetary mass of 0.46 $\pm$ 0.03 M$_{\textup{Jup}}$, a radius of ${1.05}^{+0.05}_{-0.03}$ R$_{\textup{Jup}}$, and a mean density of 0.498 $\pm$ 0.064 g cm$^{-3}$. The planet is primarily composed of hydrogen and helium, with an equilibrium temperature of $T_\textup{eq}$ = $1030^{+26}_{-19}$ K. \citet{west2009} also confirmed the discovery with SuperWASP-North and reported slightly different parameters: $M_{\rm p}$ = 0.53 $\pm$ 0.07 M$_{\textup{Jup}}$, $R_{\rm p}$ = $0.91^{+0.06}_{-0.03}$ R$_{\textup{Jup}}$, and $T_\textup{eq}$ = 960 $\pm$ 70 K. Follow-up radial velocity monitoring and adaptive optics imaging by \citet{knu2014ApJ} and \citet{ngo2015} revealed a low-mass stellar companion. The Global Architecture of Planetary Systems (GAPS) program \citep{mancini2015} measured the Rossiter–McLaughlin effect, indicating a well-aligned spin–orbit configuration.

Transit timing variations (TTVs) of WASP-11~b/HAT-P-10~b were first investigated by \citet{wang2014} through an ($O-C$) analysis. They reported a constant orbital period and found no evidence of a significant TTV signal. They also considered the possibility of an outer companion via the light-travel time (LiTE) effect \citep{irwon1952}. \citet{mancini2015} combined data from the GAPS program with the Exoplanet Transit Database (ETD) and suggested that the underestimated timing uncertainties might be due to either an unseen planetary companion or stellar activity. However, their Lomb–Scargle periodogram analysis did not reveal any significant signal, ruling out the hypothesis. Following the launch of the Transiting Exoplanet Survey Satellite \citep[TESS;][]{ricker2014}, observations of WASP-11~b/HAT-P-10~b began in 2021. \texttt{TESS} data have since been analyzed in combination with published transit timings to refine the system ephemeris, investigate orbital variations, and search for TTVs \citep{ivshina2022,maciejewski2023,Er2024}. \citet{yalcin2024} incorporated their own transit observations together with ETD data and one \texttt{TESS} sector. Their analysis showed no significant periodic changes in the TTV diagram and yielded an orbital decay rate of $dP/dE = -9.6 \pm 5.98\times10^{-10}$ day/cycle, corresponding to a stellar tidal quality factor of $Q'_{\star} > 4.1 \times 10^{2}$. In contrast, \citet{wang2024} analyzed three \texttt{TESS} sectors combined with literature mid-transit times to test for long-term orbital variations using a leave-one-out cross-validation (LOOCV) approach, and reported an increasing period derivative of $\dot{P} = 23.32 \pm 6.69$ ms yr$^{-1}$.

The study of transiting exoplanets provides not only constraints on transit timing variations (TTVs) but also valuable insights into planetary atmospheres  \citep{awiphan2016,bai2022,edwards2023}. TTV analysis is an effective method for detecting additional planets and characterizing the orbital evolution of planetary systems \citep{agol2005}. Additionally, multi-wavelength observations acquired during transit events enable the measurement of the planetary radius at different wavelengths, providing a vital baseline for future atmospheric analyses and transmission spectroscopy \citep{seager2000}. In this work, we observed transit events of WASP-11~b/HAT-P-10~b through multi-band photometry as part of the Spectroscopy and Photometry of Exoplanet Atmospheres Research Network (SPEARNET), a long-term program aimed at characterizing the atmospheres of hot transiting exoplanets using transmission spectroscopy \citep{hayes2024,athano2023}. This approach enables the simultaneous investigation of TTVs and atmospheric properties via broadband transmission spectroscopy. Together, these diagnostics provide key constraints on the formation and evolution of planetary systems.

Since 2016, SPEARNET has conducted multi-band photometric follow-up observations of WASP-11~b/HAT-P-10~b to study its transit events, TTVs, and transmission spectrum. In this paper, we present new ground-based multi-band photometric observations of 31 transits of WASP-11~b/HAT-P-10~b. These data are combined with previously published light curves and \texttt{TESS} observations to refine the planetary parameters, investigate TTVs, and constrain the optical transmission spectrum of this hot Jupiter. The structure of the paper is as follows: Section~\ref{sec:observation} describes the observational data and sources. Section~\ref{sec:LCModeling} presents the light-curve analysis. Section~\ref{sec:ttv} examines the TTVs, while Section~\ref{sec:atmosphere} discusses the optical transmission spectrum. Finally, conclusions and discussion are given in Section~\ref{sec:conclusion}.

\section{Observational Data} 
\label{sec:observation}

\subsection{SPEARNET Ground-Based Observations and Data Reduction}
\label{subsec:SPEARNETdata}

\begin{figure*}[htb]
\centering
    \includegraphics[width=0.45\textwidth]{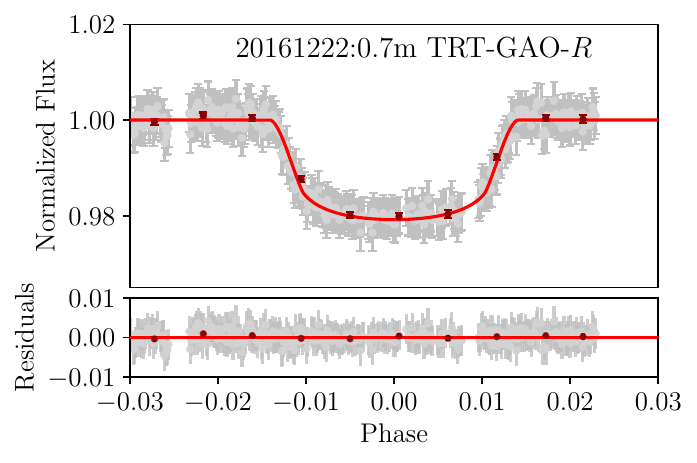}
    \caption{The normalized transit light curve of WASP-11~b/HAT-P-10~b was obtained with the 0.7 m TRT-GAO telescope in the $R$ filter on 22 December 2016, during an observation graciously conducted remotely by Her Royal Highness Princess Maha Chakri Sirindhorn while visiting the Thai National Observatory. The best-fit model from \texttt{TransitFit} is shown as a solid line, and the residuals after model subtraction are displayed in the lower panel.}
\label{fig:LCs_TGO-HRH}
\end{figure*}

From 2016 to 2024, we conducted multi-band photometric observations of WASP-11~b/HAT-P-10~b using the SPEARNET telescope network. These observations were obtained with several facilities, including the 2.4-m and 1-m Thai National Telescopes (TNT) at the Thai National Observatory (TNO) in Thailand, the 0.7-m Thai Robotic Telescope at Gao Mei Gu Observatory (TRT-GAO) in China, the 0.7-m Thai Robotic Telescope at Sierra Remote Observatories (TRT-SRO) in the USA, and the 0.7-m Regional Observatory for the Public in Nakhon Ratchasima (ROP-NM) and Chachoengsao (ROP-CC), Thailand. Instrument specifications are provided in \Cref{tab:tel}. Notably, the observation obtained on 22 December 2016 at TRT-GAO was graciously conducted remotely by Her Royal Highness Princess Maha Chakri Sirindhorn during her visit to the Thai National Observatory (\Cref{fig:LCs_TGO-HRH}). In total, we obtained 31 transit light curves, comprising 21 full and 10 partial transits, as summarized in \Cref{tab:log}.

\begin{table*}
\begin{center}
\caption{The instrumental specifications of the SPEARNET network facilities used in this work.}
\label{tab:tel}          
\small\addtolength{\tabcolsep}{-2pt}
\begin{tabular}{lcccc}
\toprule
\multirow{2}{*}{Telescope}  &  \multirow{2}{*}{CCD Camera} & CCD Pixel Size & Field Of View & Number\\
 &  & (pixels) & (arcmin$^{2}$) & of Transits\\
\hline
2.4-m TNT &   ULTRASPEC$^\dagger$ & 1024 $\times$ 1024 & 7.68 $\times$ 7.68 & 11 full, 2 partial \\
1-m TNT &  AndoriKon-M 934 & 1024 $\times$ 1024 & 23.4 $\times$ 23.4 & 1 full, 1 partial \\
0.7-m TRT-GAO & Andor iKon-L 936 & 2048 $\times$ 2048 &  20.9 $\times$ 20.9 & 2 full, 1 partial \\
0.7-m TRT-SRO &  Andor iKon-M 934 & 1024 $\times$ 1024 & 20.9 $\times$ 20.9 & 4 full, 3 partial \\
\multirow{2}{*}{0.7-m ROP-NM}  & ProLine PL16803 & \multirow{2}{*}{$4096 \times 4096$} & \multirow{2}{*}{28 $\times$ 28} & \multirow{2}{*}{2 full, 1 partial} \\
 & Monochrome & & &  \\
\multirow{2}{*}{0.7-m ROP-CC} &   ProLine PL16803 & \multirow{2}{*}{$4096 \times 4096$} & \multirow{2}{*}{28 $\times$ 28} & \multirow{2}{*}{1 full, 2 partial} \\
 & Monochrome & &  &  \\
\hline
\end{tabular}
\end{center}
{\textbf{Note}: $^\dagger$ ULTRASPEC is a high-speed frame-transfer EMCCD camera developed by \citet{dhillon2014}.}
\end{table*}

The CCD data reduction was performed using standard tasks in {\tt IRAF}\footnote{IRAF is distributed by the National Optical Astronomy Observatories, which are operated by the Association of Universities for Research in Astronomy, Inc., under a cooperative agreement with the National Science Foundation (\texttt{http://iraf.noao.edu/}).} \citep{tody1986,tody1993}. Astrometric calibration of the science images was carried out with {\tt Astrometry.net} \citep{lang2010}, and aperture photometry was performed using {\tt Source Extractor} \citep{bertin1996}. Reference stars were selected from nearby stars within 3 magnitudes of WASP-11/HAT-P-10 that showed no brightness variations. A 5$\sigma$-clipping method was applied to remove outlier points from the light curves. Differential light curves were constructed by dividing the flux of WASP-11/HAT-P-10 by the sum of the fluxes of the selected reference stars. All time stamps were converted to Barycentric Julian Date in Barycentric Dynamical Time (BJD$_\textup{TDB}$) using {\tt barycorrpy} \citep{Kano2018}. The normalized light curves are provided in machine-readable format in \Cref{tab:lightcurve}.

\begin{table*}
\begin{center}
\caption{The details of WASP-11~b/HAT-P-10~b transit observations obtained with the SPEARNET telescopes.}
\label{tab:log}          
\small\addtolength{\tabcolsep}{-2pt}
\begin{tabular}{lcccccccc}
\toprule
\multirow{2}{*}{Observation date}  & \multirow{2}{*}{Epoch$^*$} & \multirow{2}{*}{Telescope}  & \multirow{2}{*}{Filter} & \multirow{2}{*}{Exposure time (s)} & Number & Total duration of & \multirow{2}{*}{PNR (\%)$^\dagger$ } & Transit \\
  &   &   &   &  & of images & observation (hr) &   &  coverage \\
\hline
2016	Dec	7	&	394	&	0.7m	ROP-CC	&	$V$	&	60	&	218	&	4.74	&	0.37	&	Egress only		\\
			&		&	0.7m	ROP-NM	&	$R$	&	30	&	278	&	3.43	&	0.18	&	Full	\\
2016	Dec	22	&	398	&	0.7m	ROP-NM	&	$R$	&	30	&	257	&	4.52	&	0.13	&	Egress only	\\
			&		&	0.7m	TRT-GAO	&	$R$	&	30	&	356	&	5.16	&	0.12	&	Full	\\
2017	Jan	17	&	405	&	0.7m	TRT-GAO	&	$R$	&	30	&	239	&	3.19	&	0.18	&	Ingress only	\\
2017	Nov	22	&	488	&	0.7m	ROP-CC	&	$V$	&	60	&	111	&	2.74	&	0.22	&	Egress only 	\\
			&		&	0.7m	ROP-NM	&	$R$	&	30	&	188	&	4.60	&	0.24	&	Full	\\
2017	Dec	18	&	495	&	2.4-m	TNT	&	$i'$	&	3.56	&	4316	&	4.78	&	0.11	&	Full	\\
2018	Jan	2	&	499	&	2.4-m	TNT	&	$g'$	&	2.43	&	5581	&	4.43	&	0.08	&	Full	\\
2018	Nov	3	&	581	&	2.4-m	TNT	&	$r'$	&	1.95	&	9044	&	5.10	&	0.08	&	Full	\\
2018	Nov	18	&	585	&	2.4-m	TNT	&	$g'$	&	3.57	&	4868	&	5.05	&	0.07	&	Full	\\
2018	Dec	3	&	589	&	0.7m	ROP-CC	&	$V$	&	60	&	126	&	4.62	&	0.18	&	Full	\\
2020	Dec	9	&	787	&	0.7m	TRT-GAO	&	$I$	&	40	&	184	&	4.77	&	0.15	&	Full	\\
			&		&	2.4-m	TNT	&	$z'$	&	12.78	&	1704	&	6.46	&	0.10	&	Full	\\
2021	Nov	24	&	798	&	2.4-m	TNT	&	$r'$	&	5.12	&	1137	&	1.98	&	0.13	&	Ingress only	\\
2021	Dec	5	&	884	&	1-m	TNT	&	$V$	&	30	&	351	&	4.73	&	0.11	&	Full	\\
			&		&	2.4-m	TNT	&	$u'$	&	12.78	&	1243	&	4.89	&	0.62	&	Full	\\
2021	Dec	20	&	888	&	2.4-m	TNT	&	$g'$	&	12.78	&	1080	&	4.10	&	0.11	&	Full	\\
2022	Jan	30	&	899	&	2.4-m	TNT	&	$z'$	&	9.12	&	1353	&	3.81	&	0.14	&	Full	\\
2022	Dec	16	&	985	&	2.4-m	TNT	&	$u'$	&	12.78	&	765	&	3.20	&	0.53	&	Full	\\
2023	Aug	19	&	1051	&	0.7m	TRT-SRO	&	$R$	&	30	&	288	&	2.98	&	0.25	&	Egress only	\\
2023	Sep	29	&	1062	&	0.7m	TRT-SRO	&	$R$	&	30	&	373	&	3.82	&	0.27	&	Full	\\
2023	Dec	1	&	1079	&	2.4-m	TNT	&	$g'$	&	7.19	&	1230	&	2.62	&	0.10	&	Ingress only	\\
2023	Dec	5	&	1080	&	0.7m	TRT-SRO	&	$R$	&	30	&	341	&	3.53	&	0.19	&	Full	\\
2024	Jan	11	&	1090	&	2.4-m	TNT	&	$r'$	&	1.7	&	5288	&	2.63	&	0.10	&	Full	\\
2024	Aug	14	&	1148	&	0.7m	TRT-SRO	&	$R$	&	30	&	268	&	2.91	&	0.22	&	Ingress only	\\
2024	Sep	24	&	1159	&	0.7m	TRT-SRO	&	$V$	&	30	&	344	&	4.72	&	0.20	&	Ingress only	\\
2024	Oct	20	&	1166	&	0.7m	TRT-SRO	&	$I$	&	30	&	314	&	3.15	&	0.21	&	Full	\\
2024	Dec	4	&	1178	&	0.7m	TRT-SRO	&	$V$	&	30	&	364	&	3.57	&	0.19	&	Full	\\
2024	Dec	11	&	1180	&	1-m	TNT	&	$I$	&	5	&	927	&	3.09	&	0.11	&	Egress only	\\
			&		&	2.4-m	TNT	&	$z'$	&	4.21	&	3847	&	4.53	&	0.16	&	Full	\\

\hline
\end{tabular}
\end{center}
{\textbf{Note}: $^*$ Epoch = 0 is the transit on 2012 Dec 02. $^\dagger$ PNR is the photometric noise rate \citep{Fulton2011}.}
\end{table*}

\subsection{Literature Ground-based Data}
In addition to the transit light curves of WASP-11~b/HAT-P-10~b obtained from our observations, we used 10 publicly available light curves from previous studies. These include one $i'$-band and one $z'$-band light curve from the KeplerCam CCD on the FLWO 1.2 m telescope provided by \citet{bakos2009}. Three light curves from the GAPS program \citep{mancini2015} were also included: one Gunn-$r$ filter light curve observed with the Cassini 1.52 m telescope, one Cousins-$I$ filter light curve from the Zeiss 1.23 m telescope, and one Cousins-$R$ filter light curve from the IAC 80 cm telescope. Additionally, three transit light curves from \citet{maciejewski2023} were incorporated, including one $clear$ filter light curve obtained with the PIW 0.6 m Cassegrain telescope and two Cousins-$R$ light curves observed with the 1.2 m Cassegrain telescope and the 0.9 m Ritchey-Chrétien telescope, respectively. Finally, two $R$-band light curves obtained with the 1-m telescope at TÜBİTAK National Observatory from \citet{yalcin2024} were included.

\subsection{TESS Data}
WASP-11~b/HAT-P-10~b was observed by \texttt{TESS} in three sectors between 2021 and 2024, under the \texttt{TESS} Input Catalog ID TIC 85593751. Four transit light curves were obtained in Sector 42 from 2021 August 23 to September 10, seven light curves in Sector 58 from 2022 November 1 to 23, and five light curves in Sector 85 from 2024 October 31 to November 18. The \texttt{TESS} light curves were downloaded from the Mikulski Archive for Space Telescopes (MAST)\footnote{\texttt{https://archive.stsci.edu/}}. We used the Pre-Search Data Conditioning (PDC) light curves, which are calibrated by the Science Processing Operations Center (SPOC) pipeline \citep{jenkins2016}. The \texttt{TESS} timestamps were converted from Barycentric \texttt{TESS} Julian Date (BTJD) to Barycentric Julian Date (BJD$_\textup{TDB}$) by adding 2,457,000.

\section{Light-Curve Modeling}
\label{sec:LCModeling}

To derive the planetary parameters of WASP-11~b/HAT-P-10~b, we employed \texttt{TransitFit}, a Python package designed for the simultaneous fitting of multi-filter and multi-epoch exoplanet transit observations \citep{hayes2024}. This package utilizes the \texttt{batman} transit model \citep{kreidberg2015} and the \texttt{dynesty} dynamic nested sampling routines \citep{speagle2020} to estimate system parameters. The transit light curves were divided into two groups: ground-based and \texttt{TESS}. For both datasets, each individual light curve was detrended using a second-order polynomial model. This detrending was performed simultaneously with the transit light-curve fitting within \texttt{TransitFit}.

In the initial stage of the \texttt{TransitFit} retrieval, we adopted the host star's effective temperature, $T_\text{eff} = 4800 \pm 100$ K, and surface gravity, $\log g = 4.52 \pm 0.1$, as reported in the Gaia EDR3 catalogue\footnote{Gaia archive: \texttt{https://archives.esac.esa.int/gaia}}. The metallicity, [Fe/H] = $0.120 \pm 0.09$ dex, was adopted from \citet{bonomo2017}, and a circular orbit was assumed for WASP-11~b/HAT-P-10~b. Initial values for the orbital period $P$, epoch of mid-transit $T_0$ (BJD), orbital inclination $i$ (deg), semimajor axis $a$ (in units of stellar radius, $R_\ast$), and planetary radius $R_p$ (in units of $R_\ast$) for each filter are listed in \Cref{tab:initialpara}.

We first determined the best-fit orbital period, $P$, for the ground-based and \texttt{TESS} datasets. A Gaussian distribution of $3.722\,479 \pm 0.000\,001$ days was used to obtain the optimal value. The orbital inclination, $i$ (deg), and semimajor axis, $a$ (in units of stellar radius, $R_\ast$), were allowed to vary during the fitting. The best-fit values for each dataset are presented in \Cref{tab:outpara}. Final values were calculated by combining the results from both datasets using a weighted mean. We find that WASP-11~b/HAT-P-10~b has an orbital period of $3.7224797 \pm 7\times10^{-8}$ days, an inclination of $i = 88.28 \pm 0.06$ degrees, and a star–planet separation of $a/R_\ast = 12.17 \pm 0.04$. These results are consistent with previous studies within $1\sigma$.

Next, we investigated the transit timing variations (TTVs) using the \texttt{allow\_TTV} function in \texttt{TransitFit}. The final values of the orbital period, inclination, and semimajor axis were held fixed, while the mid-transit time ($t_m$) for each transit, the planetary radius ($R_p/R_\ast$), and the limb-darkening coefficients (LDCs) for each filter were allowed to vary. The derived mid-transit times ($t_m$) and corresponding epochs ($E$) are listed in \Cref{tab:midtransit}, while the values of $R_p/R_\ast$ for each filter are presented in \Cref{tab:Rp-ldc}.

The limb-darkening coefficients (LDCs) for each filter were calculated using the \texttt{Coupled} fitting mode in \texttt{TransitFit}, adopting a quadratic limb-darkening model. This calculation employed the Limb Darkening Toolkit (LDTk; \citet{husser2013,parviainen2015}) together with the host star's properties ($T_\text{eff}$, metallicity [Fe/H], and $\log g$). The derived LDCs for each filter from the \texttt{Coupled} fitting mode are presented in \Cref{tab:Rp-ldc}.

\begin{figure*}[htb]
\centering
    \includegraphics[width=0.95\textwidth]{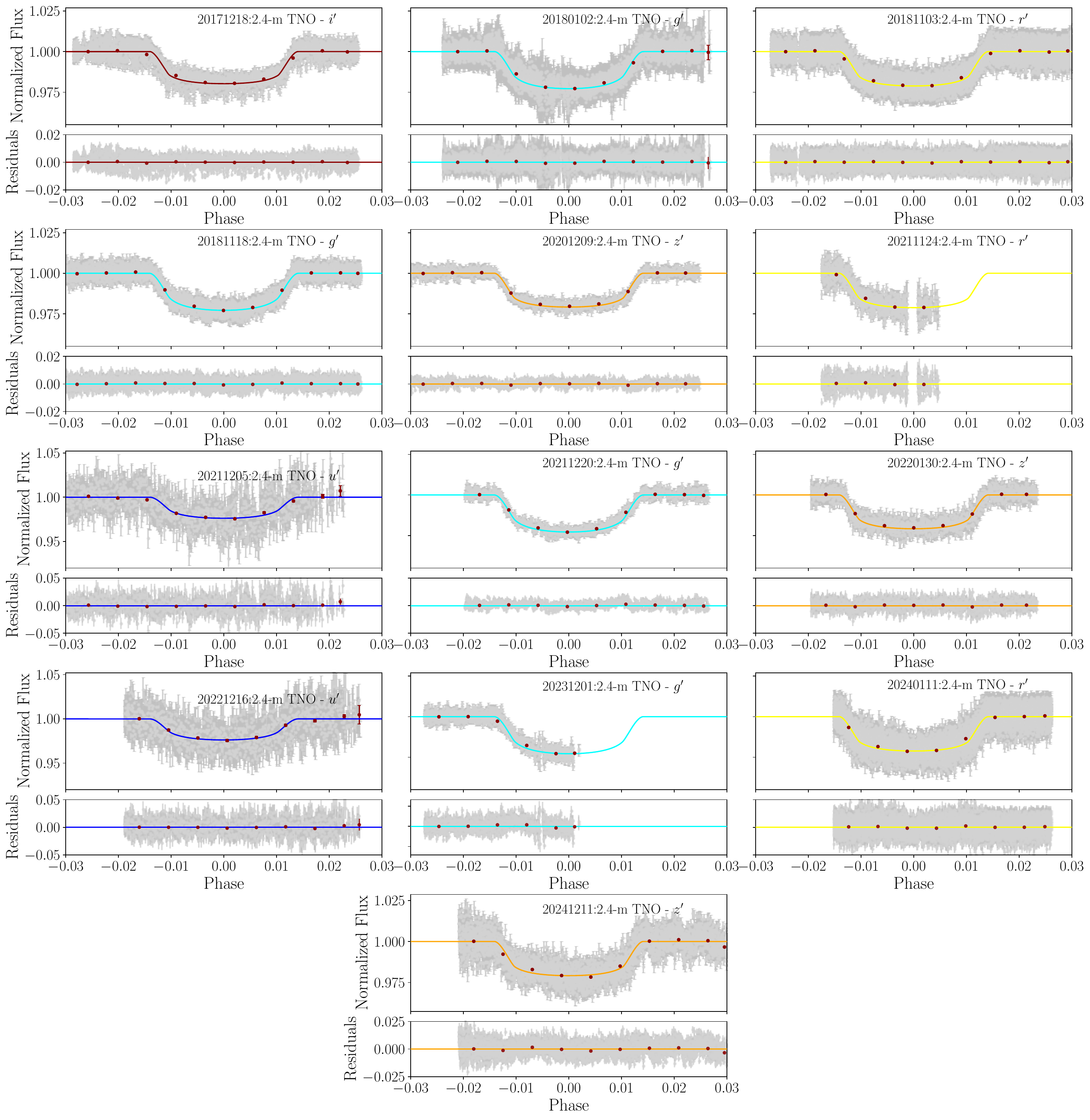}
    \caption{The normalized transit light curves of WASP-11~b/HAT-P-10~b observed with the 2.4-m telescope of the SPEARNET network (gray dots, upper panel). The best-fit model from \texttt{TransitFit} is shown as a solid line, and the residuals after model subtraction are displayed in the lower panel.}
\label{fig:LCs_TNT}
\end{figure*}

\begin{figure*}[htb]
\centering
    \includegraphics[width=0.85\textwidth]{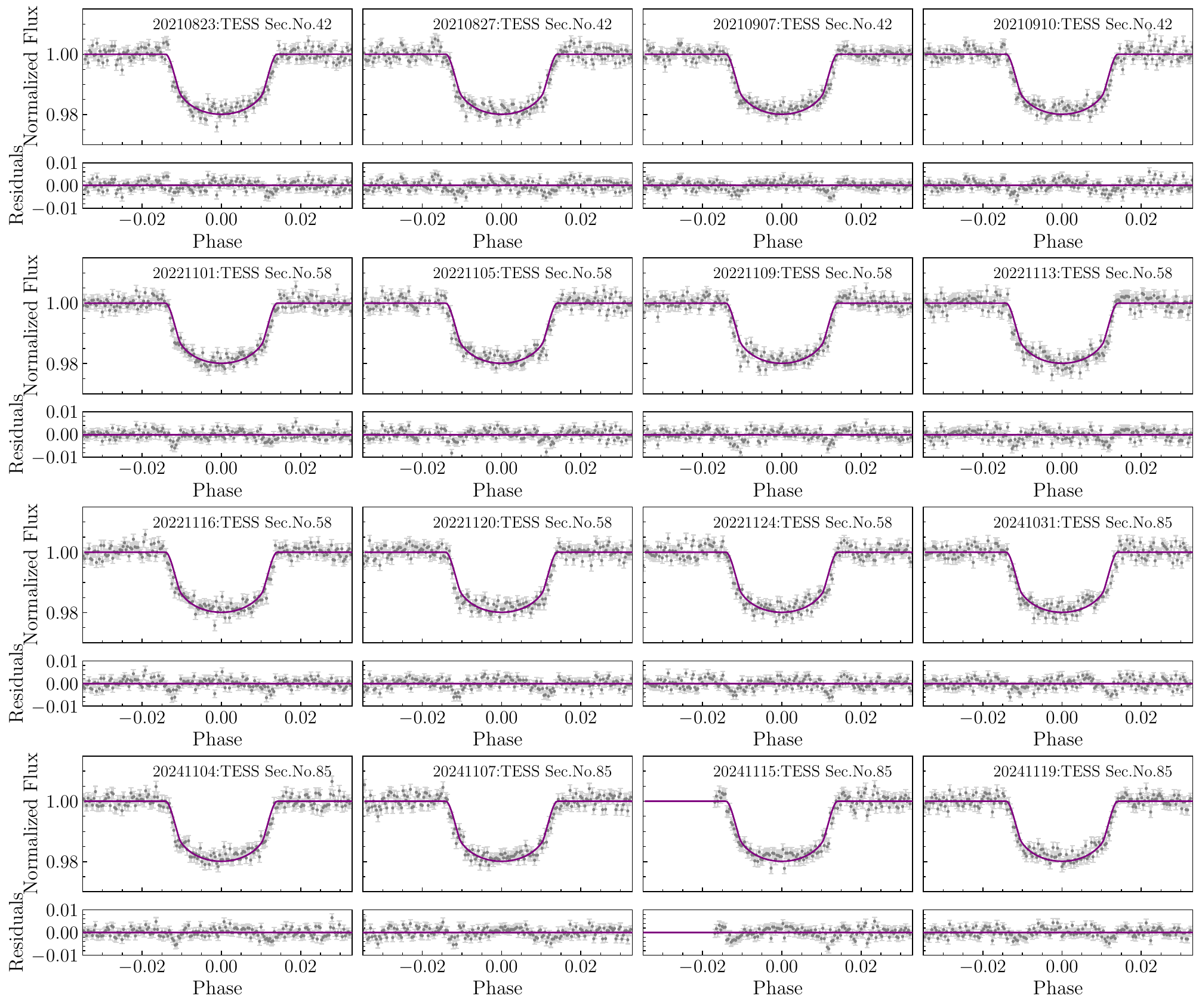}
    \caption{The normalized transit light curves of WASP-11~b/HAT-P-10~b observed by \texttt{TESS} (gray dots, upper panel), with the best-fit model from \texttt{TransitFit} shown as solid lines. The residuals after model subtraction are displayed in the lower panel.}
\label{fig:LCs_TESS}
\end{figure*}

The normalized light curves of WASP-11~b/HAT-P-10~b observed with the 2.4-m telescope and \texttt{TESS}, together with their best-fit transit models and residuals, are shown in \Cref{fig:LCs_TNT,fig:LCs_TESS}. Individual fits for the 18 light curves, obtained from the TRT-GAO, TRT-SRO, ROP-CC, ROP-NM, and 1-m TNT telescopes, are presented in \Cref{fig:LCs_TRT}.

\begin{table*}
\begin{center}
\caption{The initial parameter settings and priors for modeling planetary parameters with \texttt{TransitFit} for both ground-based and \texttt{TESS} data.}
\label{tab:initialpara}          
\small\addtolength{\tabcolsep}{-2pt}
\begin{tabular}{lcc}
\toprule
Parameter  &  Priors & Prior distribution \\
\hline
$P$ [days]         &  $3.722479 \pm 10^{-6}$  &  A Gaussian distribution  \\
${T}_{0}$ [BJD]    &  2456263.57 $\pm$ 0.002  &  A Gaussian distribution  \\
$i$ [deg]    &  (86, 90)  &  Uniform distribution  \\
{$a/R_\ast$}       &   (11, 14)    &  Uniform distribution  \\
{$R_p$/$R_\ast$}$^{*}$  &    (0.11, 0.15)    &   Uniform distribution     \\
$e$               & 0       & Fixed     \\
\hline
\end{tabular}
\end{center}
{\textbf{Notes.} The priors of $P$, ${T}_{0}$, ${\it i}$ and {a/R$_\ast$} are set as the values in \citet{bakos2009}. \\ $^*$ The same prior was used for $R_p/R_\ast$ for each filter.}
\end{table*}

\begin{table*}
\begin{center}
\caption{The best-fit planetary parameters of WASP-11~b/HAT-P-10~b from \texttt{TransitFit} compared with literature values.}
\label{tab:outpara}          
\small\addtolength{\tabcolsep}{-2pt}
\begin{tabular}{cccc}
\toprule
Reference   & $P$ (days) & $i$ (deg)  &  {$a/R_{*}$} \\
\hline
\citet{bakos2009} & 3.7224747 $\pm$ $6.5\times10^{-6}$ & $88.6^{+0.5}_{-0.4}$ &  $11.93^{+0.2}_{-0.6}$  \\
\citet{west2009} & $3.722465^{+6\times10^{-6}}_{-8\times10^{-6}}$ & $89.8^{+0.2}_{-0.8}$ & - \\
\citet{wang2014} &  $3.7224767 \pm  1.81\times10^{-6}$ &$89.14^{+0.50}_{-0.47}$& $12.27 \pm 0.24^{*} $\\
\citet{kokori2023} & $3.7224792 \pm  1.8\times10^{-7}$ & $89.1 \pm 0.5$ & $12.30 \pm 0.11$  \\
\hline
\multicolumn{4}{c}{This study}   \\
\hline
Ground-Based &  3.7224798 $\pm$ 7$\times$10$^{-8}$ & 88.2 $\pm$ 0.1 & 12.12 $\pm$ 0.04  \\
\emph{TESS} & 3.7224790 $\pm$ 2$\times$10$^{-7}$ & 89.7 $\pm$ 0.3 & 12.32 $\pm$ 0.07   \\
Weighted mean values  & 3.7224797 $\pm$ 7$\times$10$^{-8}$ & 88.28 $\pm$ 0.06 &  12.17 $\pm$ 0.04  \\
\hline
\end{tabular}
\end{center}
{\textbf{Notes.} $^*$ Reported as $0.0447 \pm 0.0002$ AU.}
\end{table*}

\begin{table*}
\begin{center}
\caption{The planet-to-star radius ($R_p/R_\ast$) and limb-darkening coefficients (LDCs) of WASP-11~b/HAT-P-10~b derived with \texttt{TransitFit} using the coupled LDCs fitting mode.}
\label{tab:Rp-ldc}
\begin{tabular}{cccccc}
\toprule
\multirow{2}{*}{Filter} & Mid-wavelength & Bandwidth & \multirow{2}{*}{ {$R_p$/$R_{*}$}} &  \multirow{2}{*}{$u_0$} & \multirow{2}{*}{$u_1$} \\ 
 & ($\mu$m) & ($\mu$m) &	   &	 &  \\ 
\hline				
$u'$-band	&	0.353	&	0.095	&	0.1369	$\pm$	0.0003	&	0.503	$\pm$	0.002	&	0.467	$\pm$	0.002	\\
$g'$-band	&	0.467	&	0.172	&	0.1363	$\pm$	0.0001	&	0.425	$\pm$	0.001	&	0.435	$\pm$	0.001	\\
$r'$-band	&	0.621	&	0.155	&	0.1342	$\pm$	0.0001	&	0.326	$\pm$	0.002	&	0.381	$\pm$	0.002	\\
$i'$-band	&	0.754	&	0.168	&	0.1298	$\pm$	0.0002	&	0.321	$\pm$	0.002	&	0.387	$\pm$	0.002	\\
$z'$-band	&	0.94	&	0.285	&	0.1331	$\pm$	0.0002	&	0.322	$\pm$	0.002	&	0.385	$\pm$	0.002	\\
$V$-band	&	0.6	&	0.24	&	0.1350	$\pm$	0.0003	&	0.325	$\pm$	0.002	&	0.382	$\pm$	0.002	\\
$I$-band	&	0.805	&	0.19	&	0.1281	$\pm$	0.0003	&	0.325	$\pm$	0.002	&	0.382	$\pm$	0.002	\\
$R$-band	&	0.672	&	0.245	&	0.1331	$\pm$	0.0001	&	0.325	$\pm$	0.002	&	0.382	$\pm$	0.002	\\
$TESS$-band	&	0.849	&	0.535	&	0.1271	$\pm$	0.0002	&	0.501	$\pm$	0.001	&	0.466	$\pm$	0.002	\\
$clear$-band	&	0.6	&	0.6	&	0.1284	$\pm$	0.0002	&	0.424	$\pm$	0.001	&	0.434	$\pm$	0.001	\\

\hline
\end{tabular}
\end{center}
\end{table*}

\section{Transit Timing Analysis }
\label{sec:ttv}
\subsection{Updated Linear Ephemeris}
By combining a long baseline of observed transit events, we used mid-transit times derived with \texttt{TransitFit} from a total of 50 events (listed in \Cref{tab:midtransit}) for the timing analysis. First, a linear ephemeris model was applied, assuming a circular orbit with a constant orbital period. The updated linear ephemeris was determined using the following relation:

\begin{equation}
t(E) = t_{0,l} + P_{\text{orb},l} \times E \ ,
\end{equation}

where $t(E)$ is the calculated mid-transit time at a given epoch $E$. The terms $t_{0,l}$ and  $P_{\text{orb},l}$ represent the reference mid-transit time and the orbital period of the linear ephemeris model, respectively. Here, $E$ denotes the epoch number, with $E=0$ defined as the transit event occurring on 2012 Dec 02 \citep{ivshina2022}.

The best-fitting parameters were determined using the \texttt{emcee} Markov Chain Monte Carlo (MCMC) package \citep{foreman2013}, employing 50 walkers and $10^{5}$ steps. To ensure convergence, a burn-in period of $10^{4}$ steps was discarded for each walker. The sampling efficiency and convergence were assessed using the mean acceptance fraction ($a_{f}$), the integrated autocorrelation time ($\tau$), and the effective number of independent samples ($N_\textup{eff}$), all of which are summarized in \Cref{tab:timing}.

From this linear ephemeris fit, the updated linear ephemeris was derived as
\begin{equation}
\label{eq:linear}
t(E) =  2456263.56879^{+0.00025}_{-0.00025} + 3.7224796^{+3\times10^{-7}}_{-3\times10^{-7}}   \ E \ .
\end{equation}

The reduced chi-squared was calculated as $\chi^{2}_{\nu} = 11.6$ with 48 degrees of freedom. The Bayesian Information Criterion (BIC) was determined to be $BIC = \chi^{2} + k \ln n = 565$, where $k$ is the number of free parameters and $n$ is the number of data points. The best-fit values for each parameter were derived from the marginalized posterior distributions shown in the corner plot. We report the median of the posterior samples as the central value, with the uncertainties corresponding to the 68\% ($1\sigma$) interval, calculated from the 16th and 84th percentiles. Using the new linear ephemeris from \Cref{eq:linear}, we constructed the $O-C$ diagram for WASP-11~b/HAT-P-10~b, which displays the timing residuals between the observed mid-transit times and the linear model, as presented in \Cref{fig:oc}.

\subsection{Searching for Evidence of Orbital Decay}
The study of close-in planets ($a < 0.1$~au), particularly hot Jupiters, is essential for investigating orbital decay, which provides a means to constrain the modified stellar tidal quality factor ($Q'_{\star}$). This phenomenon offers critical insights into the dynamical evolution of planetary systems, most notably through tidal orbital decay, mass loss, and apsidal precession. Measuring the rate of orbital decay in these systems is fundamental for characterizing stellar tidal dissipation and determining the remaining lifetimes of short-period planets \citep{macie2016, patra2017, yee2020, mannaday2020, macie2021}.

Following the study of orbital period variations in WASP-11~b/HAT-P-10~b by \citet{wang2024}, which reported an increasing orbital period in their LOOCV analysis, we examined our mid-transit times to investigate potential orbital period variations and search for signs of orbital decay in WASP-11~b/HAT-P-10~b using the following equation:

\begin{equation}
t(E) = t_{0,d} + E \times P_{orb,d} + \frac{1}{2}\frac{\textup{d} P_{d}}{\textup{d} E} E^2 \ ,
\end{equation}
where $t_{0,d}$ is a reference time of the orbital decay model. $P_{orb,d}$ is orbital period of the orbital decay model and $\textup{d}P_{d}/\textup{d}E$ is the change of orbital in each orbit.

We also performed the fitting using MCMC in the same manner as the linear model. The corner plots displaying the posterior distributions of the best-fit parameters for both models are presented in \Cref{fig:liDe_mcmc}, and the results are summarized in \Cref{tab:timing}. For the orbital decay model, we derived an orbital decay rate of $dP_{\rm d}/dE = -6^{+9}_{-9} \times 10^{-10}$ days/orbit, with a reduced chi-squared of $\chi^{2}_{\nu} = 11.1$ (47 degrees of freedom) and a BIC of 533. Using the best-fit parameters from the orbital decay model, the timing residuals as a function of epoch $E$ (calculated by subtracting the best-fit constant-period model) are shown in \Cref{fig:oc}. A comparison of the $\chi^{2}_{\nu}$ values between the linear ephemeris and orbital decay models reveals no significant differences. We therefore conclude that there is no clear evidence of orbital decay in the WASP-11/HAT-P-10 system. The derived orbital decay rate exhibits a negative trend, which is consistent with the findings of \citet{yalcin2024} but contrasts with the increasing orbital period reported by \citet{wang2024}.

In addition to the orbital decay rate obtained from our analysis, we calculated the stellar tidal quality factor, ${Q}_{\star }^{{\prime}}$, defined as \citep{gold1966}:

\begin{equation}
{Q}_{\star}^{{\prime}} = -\frac{27}{2}\pi\left (\frac{M_p}{M_\star}\right ) \left (\frac{a}{R_\star} \right )^{-5} \left (\frac{dP_d}{dE} \right)^{-1}P, 
\end{equation}
where $M_p$ and $M_{\star}$ are the masses of the planet and host star, respectively, and $a/R_\ast = 12.17 \pm 0.04$ from our fitting results. The planetary and stellar masses were adopted from \citet{bakos2009}. Using our derived orbital decay rate, we estimate $Q^\prime_\ast \sim 5.1\times10^{2}$, which is significantly lower than theoretical predictions, in the range $10^{5} \le {Q}{\star } \le 10^{7}$ \citep{penev2018}. For comparison, \citet{yalcin2024} reported no significant periodic changes in the TTV diagram and derived an orbital decay rate of $dP/dE = -9.6 \pm 5.98\times10^{-10}$ day/cycle, corresponding to a stellar tidal quality factor of $Q^\prime_\ast > 4.1 \times 10^{2}$. Our result is consistent with their lower limit and also indicates that any orbital decay is negligible and not detectable within current observational uncertainties.

\subsection{Analysis of the Apsidal Precession Model}
In addition to the linear and orbital decay analyses, the 50 mid-transit times were investigated using an apsidal precession model. This model accounts for potential inverted parabolic trends by assuming a non-zero eccentricity, $e$, and an argument of periastron, $\omega$, that precesses uniformly over time. Following \citet{gim1995}, the mid-transit times are expressed as:

\begin{equation}
t(E) = t_{0,a} + E \times P_{s} - \frac{e P_{a}}{\pi} \cos \omega(E) \ ,
\end{equation}
where
\begin{equation}
\omega(E) = \omega_{0} + \frac{d\omega}{dE} E \ ,
\end{equation}
\begin{equation}                                               
P_{s} = P_{a} \left( 1 - \frac{1}{2\pi} \frac{d\omega}{dE} \right) \ .
\end{equation}
In these equations, $t_{0,a}$ represents the reference mid-transit time, $e$ is the orbital eccentricity, $P_{a}$ is the anomalistic period, $\omega_{0}$ is the argument of periastron at $E=0$, $P_{s}$ is the sidereal period, and $d\omega/dE$ is the precession rate of the periastron.

\begin{figure*}[htb]
\centering
    \includegraphics[width=0.75\textwidth]{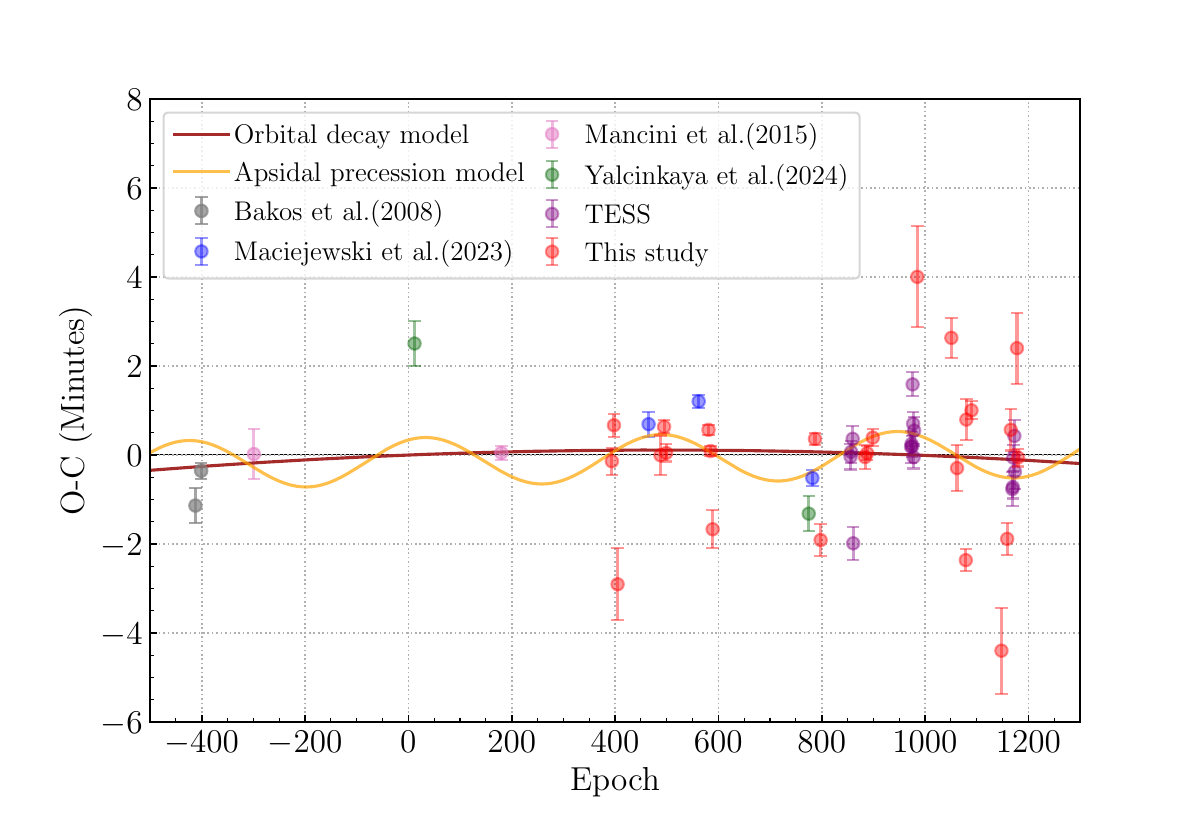}
    \caption{The $O-C$ diagram and best-fit model for WASP-11~b/HAT-P-10~b, showing data from \citet{bakos2009} (gray dots), \citet{mancini2015} (pink dots), \citet{maciejewski2023} (blue dots), \citet{yalcin2024} (green dots), \texttt{TESS} (purple dots), and this study (red dots). Timing residuals for the orbital decay and apsidal precession models are shown as brown and orange curves, respectively.}
\label{fig:oc}
\end{figure*}

The model parameters were estimated using the same MCMC configuration as the previous models. The best-fit parameters derived from the posterior distributions (\Cref{fig:liDe_mcmc}) indicate a nearly circular orbit with $e = 0.0003 \pm 0.0002$. The argument of periastron was found to be $\omega_{0} = 3^{+2}_{-1}$~rad, with a precession rate of $d\omega/dE = 0.014 \pm 0.002$~rad/orbit. This model yielded $\chi^{2}_{\nu} = 11.4$ with 45 degrees of freedom and a BIC of 531. Despite the high precession rate resulting in the sinusoidal trend shown in \Cref{fig:oc}, the $\chi^{2}_{\nu}$ values remain consistent with those of the linear and orbital decay models, showing no statistically significant improvement.

Comparing the BIC values among the three models, the apsidal precession model yielded the lowest BIC of 531, followed by the orbital decay model with a BIC of 533, and the linear model with a BIC of 565. The difference between the apsidal and decay models is statistically small with a $\Delta BIC$ of 2, providing no decisive evidence for one over the other. However, our analysis shows that the orbital decay rate lacks statistical significance, and the derived tidal quality factor for the host star is three orders of magnitude smaller than theoretical predictions. Given these discrepancies and the lack of strong statistical support, the orbital decay model can be ruled out based on the present timing data.

\begin{table*}
\begin{center}
\caption {The uniform priors and best-fit parameters from MCMC transit timing analyses.}
\label{tab:timing}
\begin{tabular}{lcc}
\hline
\hline
Parameter & Uniform distribution priors & Best fit values\\
\hline
Constant-period Model  &      &       \\
$P_{orb,l}$ [days]               &  (3.72246, 3.72249)    & $3.7224796^{+3\times10^{-7}}_{-3\times10^{-7}}$      \\
$t_{0,l}$ [BJD$_{\textup{TDB}}$ + 2450000]         & (6263.567, 6263.574)    & $6263.56879^{+0.0003}_{-0.0003}$    \\
${\chi}^{2}_{red}$ &   \multicolumn{2}{c}{11.6}          \\
BIC & \multicolumn{2}{c}{565} \\
$a_{f}$ & \multicolumn{2}{c}{0.6} \\
$\tau$  & \multicolumn{2}{c}{40}  \\
$N_\textup{eff}$ & \multicolumn{2}{c}{122606}  \\
\hline
Orbital Decay Model    &        &	    \\
$P_{orb,d}$ [days]     &  (3.72246, 3.72249)    & $3.7224799^{+6\times10^{-7}}_{-6\times10^{-7}}$      \\
$t_{0,d}$ [BJD$_{\textup{TDB}}$ + 2450000]          &  (6263.567, 6263.574)    &  $6263.56879^{+0.0002}_{-0.0002}$      \\
 $dP_{\rm d}/dE$ [days/orbit]        & (-0.5, 0.5)     &  $-6^{+9}_{-9}\times 10^{-10}$
    \\
${\chi}^{2}_{red}$ &   \multicolumn{2}{c}{11.1}          \\
BIC &  \multicolumn{2}{c}{533} \\
$a_{f}$ & \multicolumn{2}{c}{0.6} \\
$\tau$  & \multicolumn{2}{c}{49}  \\
$N_\textup{eff}$ & \multicolumn{2}{c}{101544}  \\
\hline
Apsidal Precession Model   &        &	    \\
$P_{s}$ [days]                   &  (3.72246, 3.72249)      &  $3.7224797^{+3\times10^{-7}}_{-3\times10^{-7}}$      \\
$t_{0,a}$ [BJD$_{\textup{TDB}}$ + 2450000]          &  (6263.567, 6263.574)  & $6263.56869^{+0.0003}_{-0.0003}$   \\
$e$                              &  (0, 0.002)         &   $0.0003^{+0.0002}_{-0.0002}$   \\
$\omega_{0}$ [rad]               &  (0, 2$\pi$)    &   $3 ^{+2}_{-1}$        \\
$d\omega/dE$ [rad/orbit]  &  (0, 0.02)         &  $0.014^{+0.002}_{-0.002}$    \\
${\chi}^{2}_{red}$ &   \multicolumn{2}{c}{11.4}          \\
BIC &  \multicolumn{2}{c}{531} \\
$a_{f}$ & \multicolumn{2}{c}{0.4} \\
$\tau$  & \multicolumn{2}{c}{1506}  \\
$N_\textup{eff}$ & \multicolumn{2}{c}{3320}  \\
\hline
\end{tabular}\\
\end{center}           
\end{table*}

\subsection{Line-of-Sight Acceleration}
Furthermore, orbital decay can be investigated by measuring line-of-sight acceleration, a phenomenon known as the Rømer effect. This effect shows that if the center of mass of the star-planet system accelerates along the line of sight with a magnitude of $\dot{\nu}_\textup{RV}$, the observed orbital period would change \citep{yee2020,bouma2020,macie2021,mannaday2022}. Based on this phenomenon, an acceleration toward the observer causes the period to decrease, while an acceleration away from the observer causes the period to increase. Following the formula from \citet{macie2021}, the relationship between the period change and the acceleration is given by:
\begin{equation}
\label{eq:vdot}
    \dot{\nu}_{RV} = \frac{\dot{P_q}}{P_q}c \, 
\end{equation}
where $P_q$ represents the orbital period from the orbital decay model ($P_{orb,d}$in \Cref{tab:midtransit}), $\dot{P_q}$ is the period derivative, and $c$ is the speed of light. By applying the formula $\dot{P_q} = \frac{1}{P_q}\frac{dP_d}{dE}$, we found that $\dot{P_q}$= ${-5.5}^{+7.6}_{-7.6}$ ms yr$^{-1}$. From \Cref{eq:vdot}, the radial velocity acceleration ($\dot{\nu}_\textup{RV}$) is derived to be $-0.014 \pm 0.019 \text{ m s}^{-1} \text{ d}^{-1}$. To confirm this value, we calculated the linear acceleration from RV observations ($\dot{\gamma} = \dot{\nu}_{\text{RV}}$) by using the \texttt{RadVel} Python package to fit the radial velocity curves \citep{fulton2018}. We utilized the RV data from \citet{west2009}. To perform the RV fitting, the orbit was assumed to be circular ($e=0$). The orbital period and mid-transit time were constrained using Gaussian priors from the values in \Cref{tab:initialpara}. The argument of periastron ($\omega$) was fixed to zero, while the RV semi-amplitude ($K$), the center-of-mass velocity ($\gamma$), the linear RV trend ($\dot{\gamma}$), the quadratic RV trend ($\ddot{\gamma}$) and the "jitter" radial velocity were allowed to vary. From the fitting, we obtained $K= 82.04 \pm 1.08$ m s$^{-1}$ and $\gamma = 4.914 \pm 0.008$ km s$^{-1}$, which are consistent with the results reported by \citet{west2009}. However, the fitted linear RV trend yields a positive value of $\dot{\gamma} = (1 \pm 50) \times 10^{-5}$ m s$^{-1}$ d$^{-1}$. This discrepancy value differs by approximately three orders of magnitude from the estimate derived from \Cref{eq:vdot}. Furthermore, the large uncertainties in both methods, based on the timing data ($\dot{\nu}_\textup{RV}$) and the RV fitting ($\dot{\gamma}$), make it difficult to draw clear conclusions about this phenomenon. Therefore, a longer observational baseline from both transit timing and radial velocity data is needed to confirm this effect in the system.

\subsection{The analysis of TTVs periodicity}
\label{sec:periodical}
The search for unseen planetary companions using Lomb–Scargle periodogram analysis \citep{lomb1976} was initially conducted for this system by \citet{mancini2015} and later extended in studies by \citet{maciejewski2023} and \citet{Er2024}. In the present study, we investigated whether variations in the orbital period could be caused by additional planets influencing the system. The timing residuals ($O-C$) from \Cref{tab:midtransit} were used to search for periodic TTV signals.  These signals were analyzed using the Generalized Lomb–Scargle periodogram \citep[GLS;][]{zech2009}. The False Alarm Probability (FAP) for the highest power peaks was determined using the analytical probability method described by \citet{zech2009}, as implemented in the \texttt{PyAstronomy} package \citep{pya2019}.\footnote{PyAstronomy: \texttt{https://github.com/sczesla/PyAstronomy}}

\begin{figure*}[htb]
\centering
    \includegraphics[width=0.46\textwidth,trim ={0.5cm 0 0 0.3cm}]{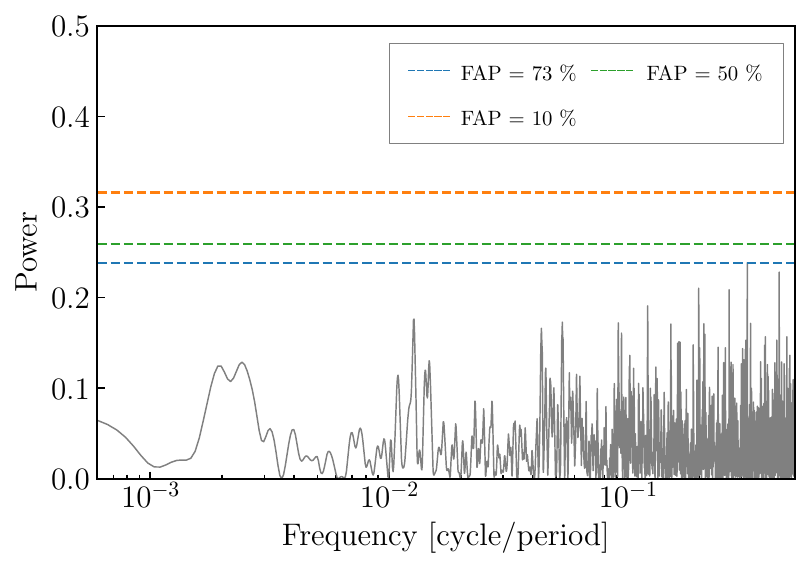} 
\caption{The GLS periodogram of the timing residuals from 50 mid-transit times derived with \texttt{TransitFit}. The false-alarm probability (FAP) levels are indicated by dashed lines.}
\label{fig:gls}
\end{figure*}

The GLS results are shown in \Cref{fig:gls}. The periodogram shows the highest-power peak at 0.24, corresponding to a frequency of 0.3164 $\pm$ 0.0002 cycles per period and a false-alarm probability (FAP) of 73\%. Based on this analysis, no statistically significant TTV signal is detected that would indicate the presence of an additional planet in the WASP-11/HAT-P-10 system.

\section{Atmospheric Modeling in Optical Wavelength} 
\label{sec:atmosphere}
Using transmission spectroscopy, the atmospheric composition of a planet can be investigated in detail \citep{seager2000}. For hot Jupiters, numerous studies have explored the presence of atomic and molecular species such as Na, K, and TiO/VO, as well as the effects of clouds, hazes, and Rayleigh scattering in the optical wavelength range \citep{sing2016,spy2023,fairman2024}. Considering the {$R_p/R_\ast$} values derived from different filter bands with \texttt{TransitFit} (listed in \Cref{tab:Rp-ldc}, Section~\ref{sec:LCModeling}), we find that the planetary radius appears larger in the blue band, which may be indicative of Rayleigh scattering \citep{leca2008,kirk2017}.

We performed the atmospheric retrieval analysis using the open-source code PLanetary Atmospheric Transmission for Observer Noobs (\texttt{PLATON}\footnote{PLATON: \texttt{https://github.com/ideasrule/platon}} \citet{zhang2019}). \texttt{PLATON} provides a fast framework for forward modeling and retrieval of exoplanet atmospheres. It employs \texttt{PyMultiNest}, a nested sampling algorithm, to compute the Bayesian evidence ($\mathcal{Z}$) and posterior distributions for the retrieval analysis \citep{feroz2009}. For this study, the planet-to-star radius ratios ($R_p/R_{*}$) listed in \Cref{tab:Rp-ldc}, covering wavelengths from 0.3 to 0.9 $\mu$m across all filters except the $clear$ filter, were used for atmospheric retrieval.

During the retrieval, the forward optical transmission spectrum was modeled assuming an isothermal atmosphere in chemical equilibrium, with the cloud-top pressure set to $10^{5}$~Pa. The host star radius was adopted from \citet{bakos2009}. We retrieved the planetary temperature ($T$), the metallicity ($\log Z$), and the carbon-to-oxygen ratio (C/O), which determines the relative molecular abundances. To account for clouds and hazes, we included a scattering factor and a scattering slope. An error multiplier was incorporated to scale the observational uncertainties. To ensure a robust exploration of the parameter space, the fitting was performed with 1,000 live points and a termination criterion of $\Delta \ln Z < 0.1$ using the nested sampling method. This process generated posterior distributions for each parameter, with the resulting corner plots displaying the median values and the associated $1\sigma$ intervals based on the 16th and 84th percentiles. The priors and the retrieved results are summarized in \Cref{tab:atmosphere}.

\begin{figure*}[htb]
\centering
    \includegraphics[width=0.8\textwidth,trim ={0.5cm 0 0 0.3cm}]{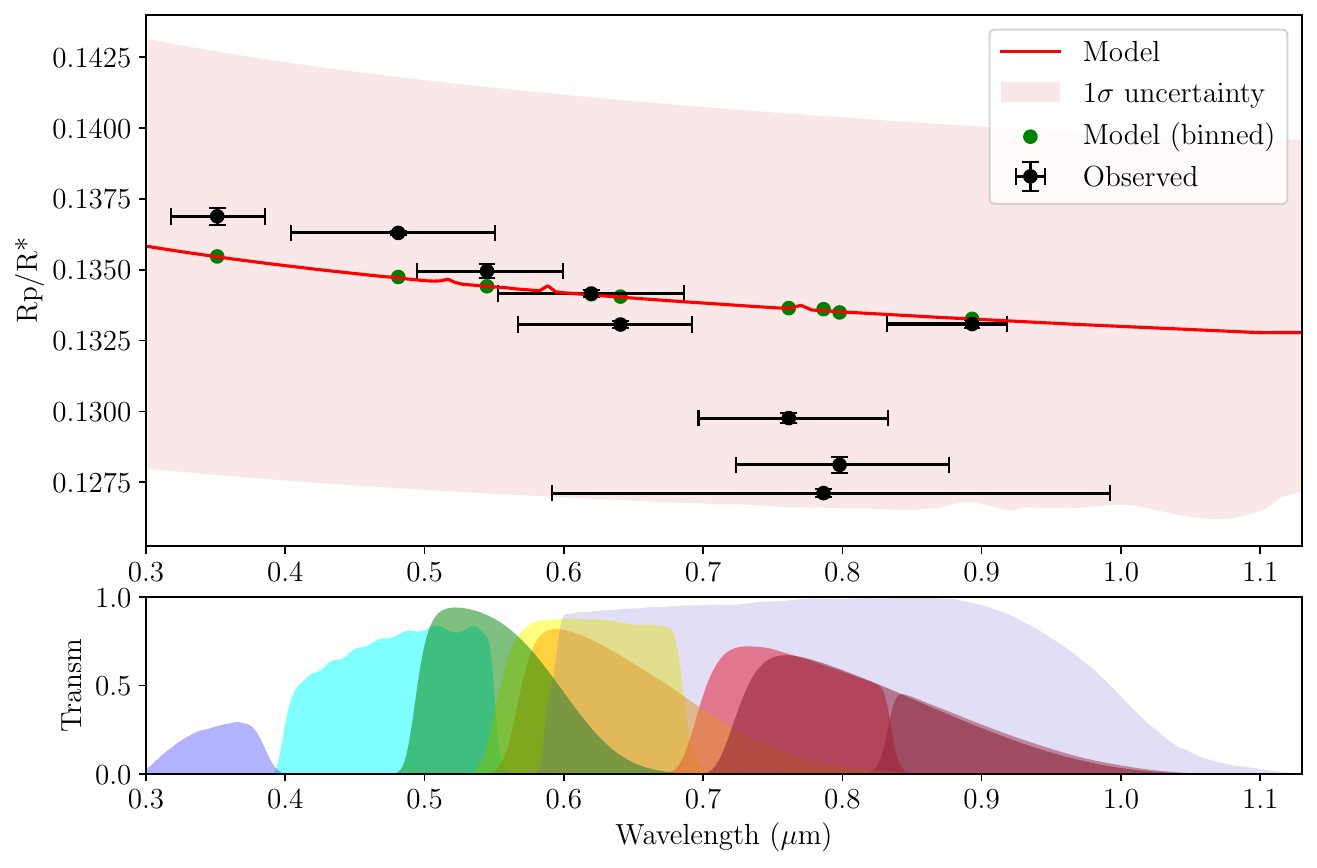} 
\caption{Top panel: Retrieved transmission spectrum of WASP-11~b/HAT-P-10~b with forward models generated using \texttt{PLATON}.
Bottom panel: Band-pass filter profiles for $u'$, $g'$, $V$, $r'$, $R$, $i'$, $I$, $TESS$, and $z'$ (from left to right).}
\label{fig:spectrum}
\end{figure*}

The atmospheric retrieval indicates that WASP-11~b/HAT-P-10~b has an isothermal temperature of approximately 1000~K. At a cloud-top pressure of $10^{5}$~Pa, the planetary radius is $0.99 \pm 0.02$~$R_{\rm Jup}$, with a corresponding mass of $0.49 \pm 0.01$~$M_{\rm Jup}$. The host star radius was found to be $0.79 \pm 0.01$~$R_\odot$. These results are summarized in \Cref{tab:atmosphere} and \Cref{fig:spectrum}, with the posterior distributions of the model parameters presented in \Cref{fig:contour-spec}. A strong Rayleigh scattering slope is observed from blue optical to near-infrared wavelengths, characterized by a scattering factor of $\log f_{\rm scat} = 3 \pm 1$ and a C/O ratio of $1.1 \pm 0.6$. Due to the large uncertainty in the C/O ratio, the molecular composition of the atmosphere cannot be robustly constrained. Consequently, additional near-infrared observations are necessary to better characterize the atmospheric constituents of WASP-11~b/HAT-P-10~b.

The strong Rayleigh scattering observed in the atmosphere of WASP-11~b/HAT-P-10~b is similar to that detected in other planetary systems, such as WASP-6~b \citep{nikolov2015,carter2020,grubel2025} and HAT-P-12b \citep{sing2016,wong2020}, where high-altitude Rayleigh-scattering clouds contribute to the effect. Since the host star of WASP-11~b/HAT-P-10~b is a K-type star in a binary system, the observed scattering signature could potentially be influenced by stellar activity or contamination from the companion star, as discussed by \citet{jiang2021}. However, no clear signs of stellar activity are evident in the \texttt{TESS} light curves. Therefore, long-term monitoring of the transmission spectrum at higher spectral resolution is required to confirm the presence and origin of the strong Rayleigh scattering in this system.

\begin{table*}
\begin{center}
\caption{Prior ranges, distribution types, and the resulting best-fit parameters obtained from the \texttt{PLATON} atmospheric retrieval for WASP-11~b/HAT-P-10~b. The retrieved values represent the medians of the posterior distributions, with uncertainties corresponding to the $1\sigma$ intervals.}
\label{tab:atmosphere}          
\small\addtolength{\tabcolsep}{-2pt}
\begin{tabular}{lccc}
\toprule
Parameter   & Prior range &  Prior Distribution & Retrieved Values   \\
\hline
$R_s$ [$R_\odot$]   &  0.79 $\pm$ 0.01 &   Gaussian  & $0.79^{+0.01}_{-0.01}$   \\
$M_p$ [$M_\textup{jup}$]   &  0.487 $\pm$ 0.01  &  Gaussian  & $0.49^{+0.01}_{-0.01}$  \\
$R_p$ [$R_\textup{jup}$]   &  (0.804, 1.206)  & Uniform  &  $0.99^{+0.02}_{-0.02}$   \\
$T$ [$K$]     &  (450, 1350)  & Uniform  & $1000^{+200}_{-300}$     \\
$\log_{Z/Z_{\odot}}$  &  (-0.5, 2.0)  & Uniform  &  $0.5^{+0.8}_{-0.7}$   \\
$\log_{\textup{Scattering Factor}}$  &  (0, 5)  & Uniform  &  $3^{+1}_{-1}$   \\
C/O ratio        &   (0.1, 2.0)  & Uniform  &   $1.1^{+0.6}_{-0.6}$   \\
Error multiple   &  (0.5, 20)  & Uniform  &  $18^{+2}_{-2}$   \\
\hline
\end{tabular}
\end{center}
{\textbf{Notes.} The priors of $R_s$, $M_p$ and $R_p$ are set as the values from \citet{bakos2009}.}
\end{table*}

\section{Summary and Conclusions}
\label{sec:conclusion}
 
We present a new set of 31 optical transit light curves of the hot Jupiter WASP-11~b/HAT-P-10~b obtained with the ground-based SPEARNET telescope network. This new dataset was combined with previously published ground-based light curves and \texttt{TESS} data. All light curves were modeled using \texttt{TransitFit}. From the fitting, WASP-11~b/HAT-P-10~b exhibits an orbital period of $3.7224797 \pm 7\times10^{-8}$ days, an inclination of $i = 88.28 \pm 0.06^\circ$, and a star–planet separation of $a/R_\ast = 12.17 \pm 0.04$, all consistent with previous studies.

Mid-transit times from a total of 50 epochs, derived using \texttt{TransitFit}, were used to update the linear ephemeris and to search for evidence of orbital decay and apsidal precession. The updated linear ephemeris is $t(E) = 2456263.56879 \pm 0.00025 + (3.7224796 \pm 3\times10^{-7}) \times E$. While the model comparisons show similar reduced chi-squared and BIC values, the derived tidal quality factor is three orders of magnitude smaller than theoretical expectations. The orbital decay scenario is physically inconsistent and is ruled out for the current timing data. Additionally, the line-of-sight acceleration analysis finds a discrepancy between the acceleration values derived from TTV and RV data. Therefore, this study and the current data do not support the presence of this phenomenon. We also searched for transit timing variation (TTV) signals from potential unseen planetary companions using periodogram analysis. However, no significant TTV signals were detected due to the high false alarm probability, indicating the absence of additional planets within detectable limits. We acknowledge that the reduced chi-squared values for both the linear and orbital decay models are relatively high. We attribute this primarily to the sensitivity of high-precision observations to low-level astrophysical noise, such as stellar activity, starspot crossings, or granulation. These physical phenomena introduce small fluctuations in mid-transit times. Furthermore, the residuals likely contain low-amplitude dynamical variations or remaining differences between datasets that are not fully captured by simple orbital models. Since our periodogram analysis did not identify a definitive periodic signal, we conclude that the linear model remains the best choice. The high reduced chi-squared reflects the difficulty of combining data from multiple telescopes when stellar and dynamical noise are present over a long period.

For atmospheric characterization, the planet-to-star radius ratios ($R_p/R_*$) from nine filters, spanning optical to near-infrared wavelengths, were used in transmission spectroscopy analysis with \texttt{PLATON}. The results reveal a strong Rayleigh scattering slope from the blue optical to near-infrared, with a C/O ratio of $1.1^{+0.6}_{-0.6}$. This strong Rayleigh scattering is similar to that seen in WASP-6b \citep{nikolov2015,carter2020,grubel2025} and HAT-P-12b \citep{sing2016,wong2020}. Although the host star is a K-type star in a binary system, no clear stellar activity is evident in the \texttt{TESS} light curves. Long-term, high-resolution spectroscopic monitoring is therefore needed to confirm the presence and origin of the Rayleigh scattering, making WASP-11~b/HAT-P-10~b a promising target for future observations.

\vspace{5mm}
%\begin{acknowledgments}
We thank the referee for their comments and suggestions, which have improved the quality of this work. This work is supported by the Fundamental Fund of Thailand Science Research and Innovation (TSRI) through the National Astronomical Research Institute of Thailand (Public Organization) (FFB680072/0269). Ing-Guey Jiang acknowledges support from the National Science and Technology Council (NSTC), Taiwan, under grants NSTC 113-2112-M-007-030 and NSTC 114-2112-M-007-029. This paper is based on observations made with ULTRASPEC at the Thai National Observatory, the Thai Robotic Telescopes, and the Regional Observatories for the Public under the operation of the National Astronomical Research Institute of Thailand (Public Organization). This work used the available data based on observations made with the \emph{TESS} mission, obtained from the MAST data archive at the STScI \citep{10.17909/t9-nmc8-f686}. Funding for the \emph{TESS} mission is provided by the NASA Explorer Program. STScI is operated by the AURA, Inc., under NASA contract NAS 5–26555.
%\end{acknowledgments}


\vspace{5mm}

\facilities{TESS, 2.4-m (TNT), 1-m (TNT), 0.7-m (TRT-GAO), 0.7-m (TRT-SRO), 0.7-m (ROP-NM) and 0.7-m (ROP-CC)}

\software{\texttt{sextractor} \citep{bertin1996}, \texttt{Astrometry.net} \citep{lang2010}, \texttt{TransitFit} \citep{hayes2024}, \texttt{PLATON} \citep{zhang2019}.}


%% Figure and Table~counter will not reset.

\appendix
\section{WASP-11~b/HAT-P-10~b Transit Light Curves from SPEARNET.}
The transit light curves of WASP-11~b/HAT-P-10~b, obtained from SPEARNET ground-based multi-band photometric observations, are provided in \Cref{tab:lightcurve}.
\begin{table*}
\begin{center}
\caption{The normalized transit light curves of WASP-11~b/HAT-P-10~b, observed by the SPEARNET telescope network.}
\label{tab:lightcurve}          
\small\addtolength{\tabcolsep}{-2pt}
\begin{tabular}{lcccc}
\toprule
\multirow{2}{*}{Epoch} & \multirow{2}{*}{BJD} & \multirow{2}{*}{Normalized Flux} & Normalized flux   \\
      &     &                 & Error       \\
\hline
394	&	2457730.12572	&	1.009	&	0.005	\\
	&	2457730.12798	&	0.990	&	0.006	\\
	&	2457730.12947	&	1.002	&	0.007	\\
	&	...	&		&	...	\\
\hline							
398	&	2457745.03460	&	1.004	&	0.004	\\
	&	2457745.03500	&	0.999	&	0.004	\\
	&	2457745.03539	&	1.000	&	0.004	\\
	&	...	&		&	...	\\
\hline							
405	&	2457771.07131	&	0.995	&	0.004	\\
	&	2457771.07174	&	0.998	&	0.004	\\
	&	2457771.07217	&	0.999	&	0.004	\\
	&	...	&		&	...	\\
\hline
... & ...   & ...    & ...        \\
\hline
\end{tabular}
\end{center}
{\textbf{Note}: The full Table~is available in machine-readable form.}
\end{table*}

\section{Mid-transit Times and $O-C$ of WASP-11~b/HAT-P-10~b}
The mid-transit times derived with \texttt{TransitFit} and the timing residuals ($O-C$), calculated from \Cref{eq:linear}, are listed in \Cref{tab:midtransit}.
\begin{table*}
\begin{center}
\caption {Mid-transit times ($t_{m}$) and timing residuals ($O-C$) for 50 epochs of WASP-11~b/HAT-P-10~b.}
\label{tab:midtransit}
\begin{tabular}{lccc}
\toprule
\multirow{2}{*}{Epoch} & $t_{m} +2450000$ & $(O-C)$ &   \multirow{2}{*}{Ref} \\
                       & (BJD$_{\textup{TDB}}$)    &   (days)      &     \\
\hline
-412	&	4729.90642	$\pm$	0.00027	&	-0.00079	&	(a)	\\
-401	&	4770.85432	$\pm$	0.00012	&	-0.00025	&	(a)	\\
-299	&	5150.54741	$\pm$	0.00039	&	0.00001	&	(b)	\\
...	&	...	&	...	&	...	\\
...	&	...	&	...	&	...	\\

\hline
\end{tabular}
\end{center}
{\textbf{Notes.} The full Table~is available in machine-readable form. Data Source: (a) \citet{bakos2009} (b) \citet{mancini2015} (c) \citet{yalcin2024}, (d) This study, (e) \citet{maciejewski2023} and (f) \texttt{TESS}. \\ }
\end{table*}

\counterwithin{figure}{section}
\section{Individual WASP-11~b/HAT-P-10~b transit light curves from SPEARNET observations.}
\begin{figure*}[htb]
\begin{tabular}{cc}
    \includegraphics[scale=0.42]{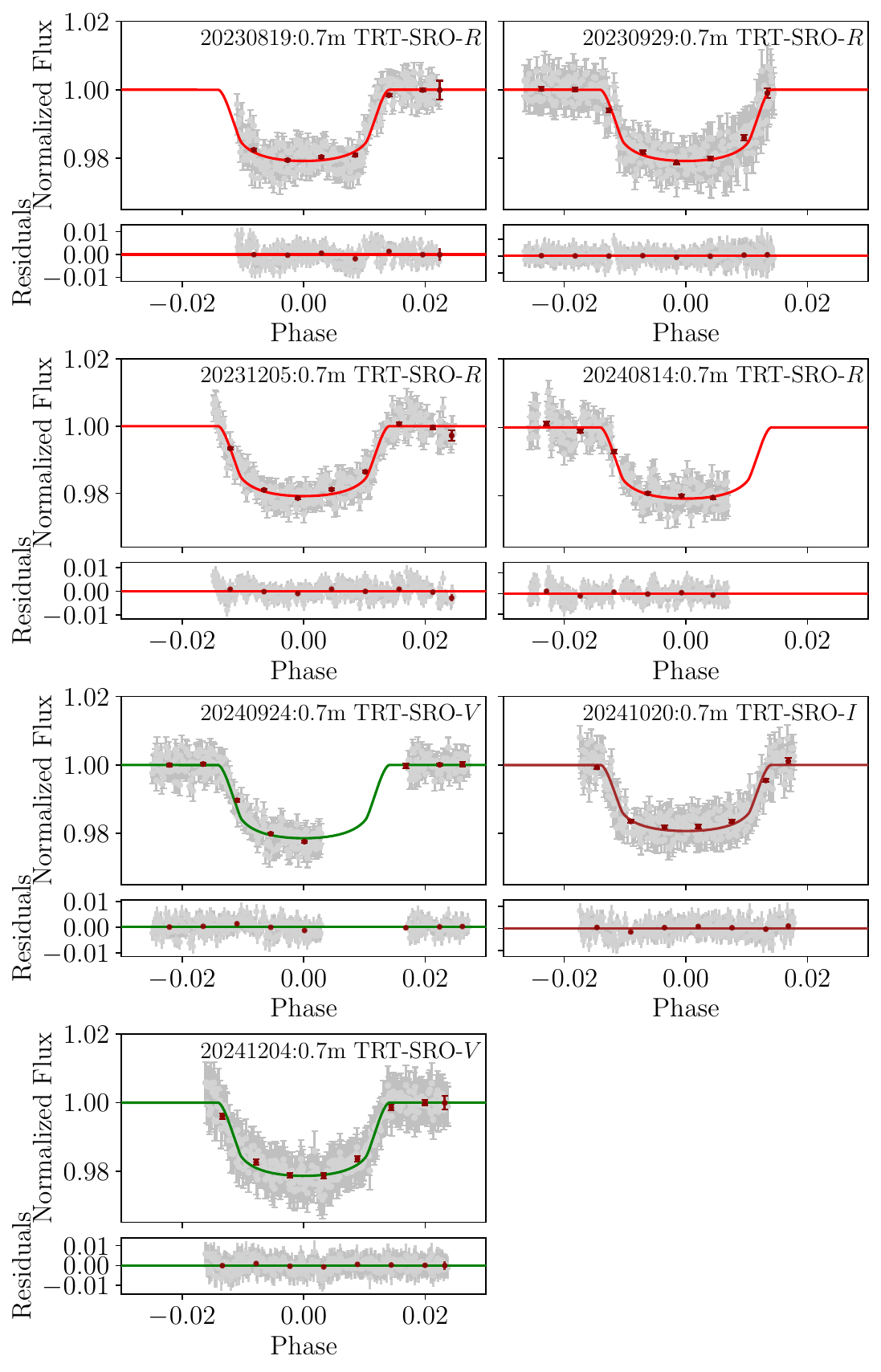} 
    \includegraphics[scale=0.42]{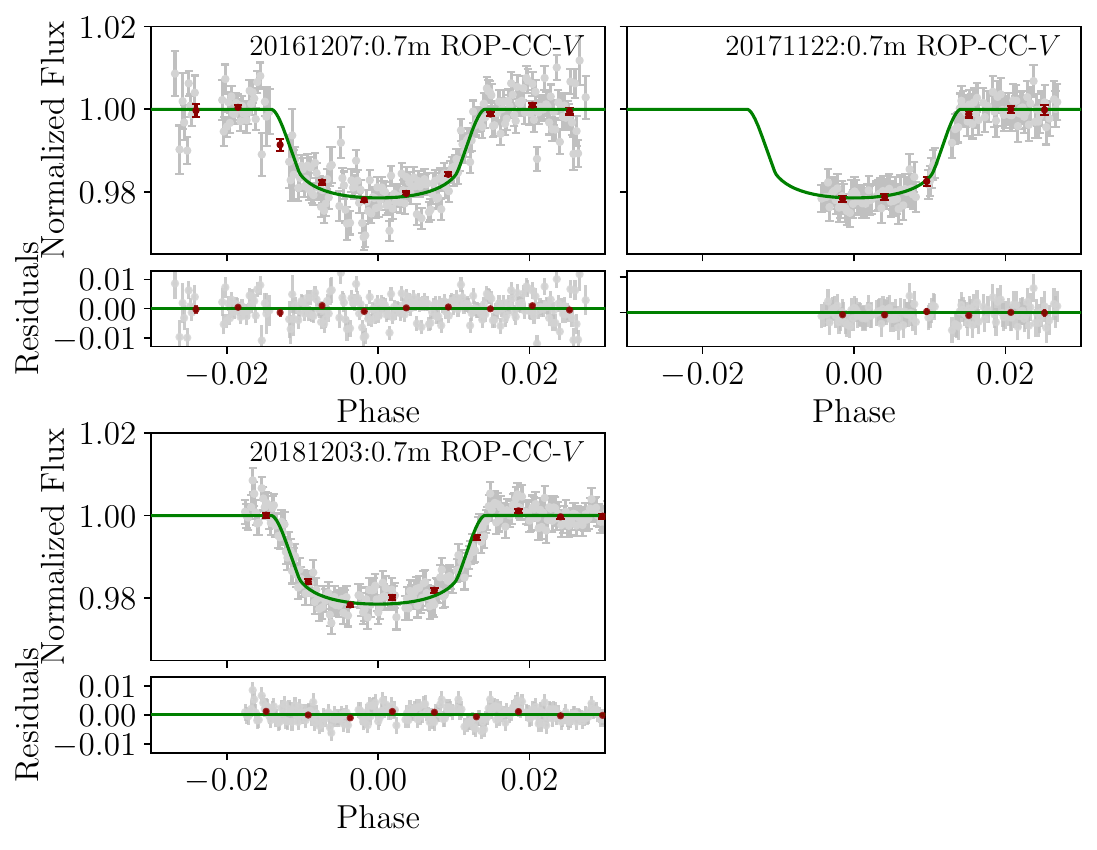} \\  
    \includegraphics[scale=0.42]{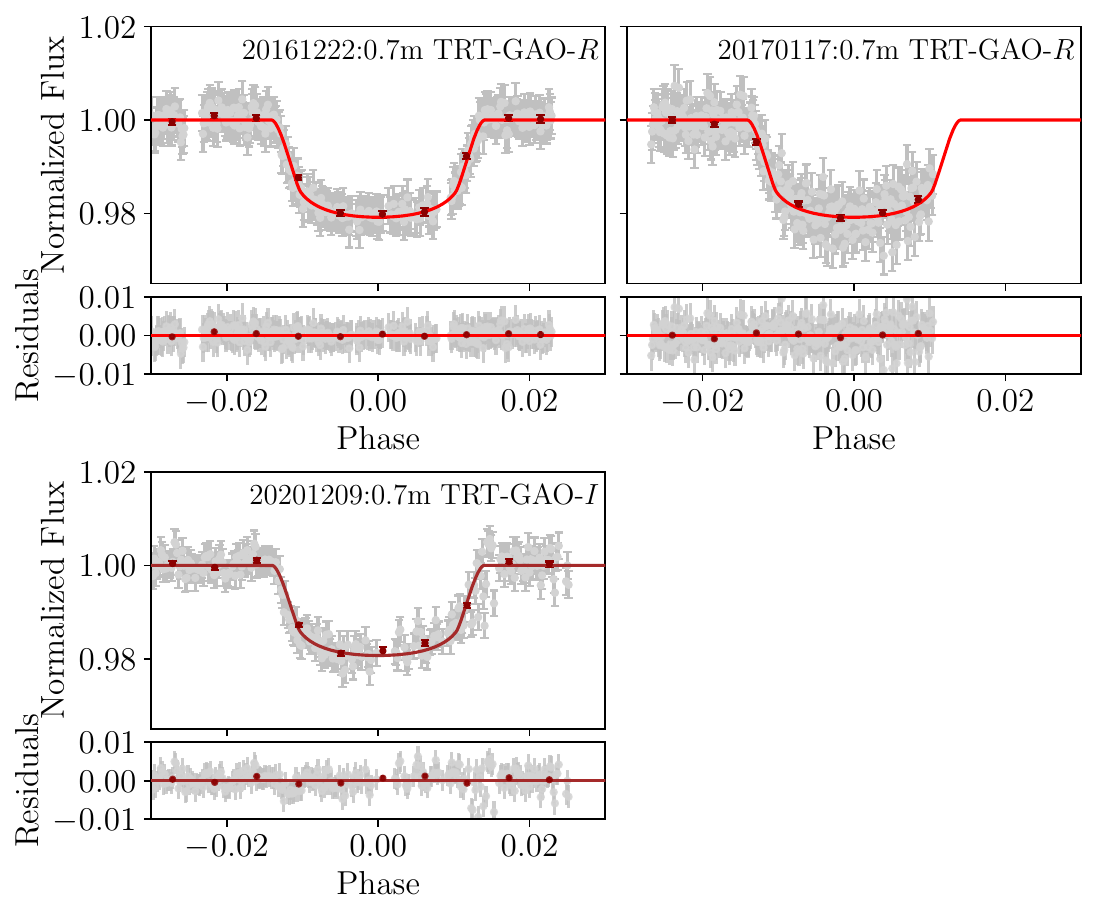}
    \includegraphics[scale=0.42]{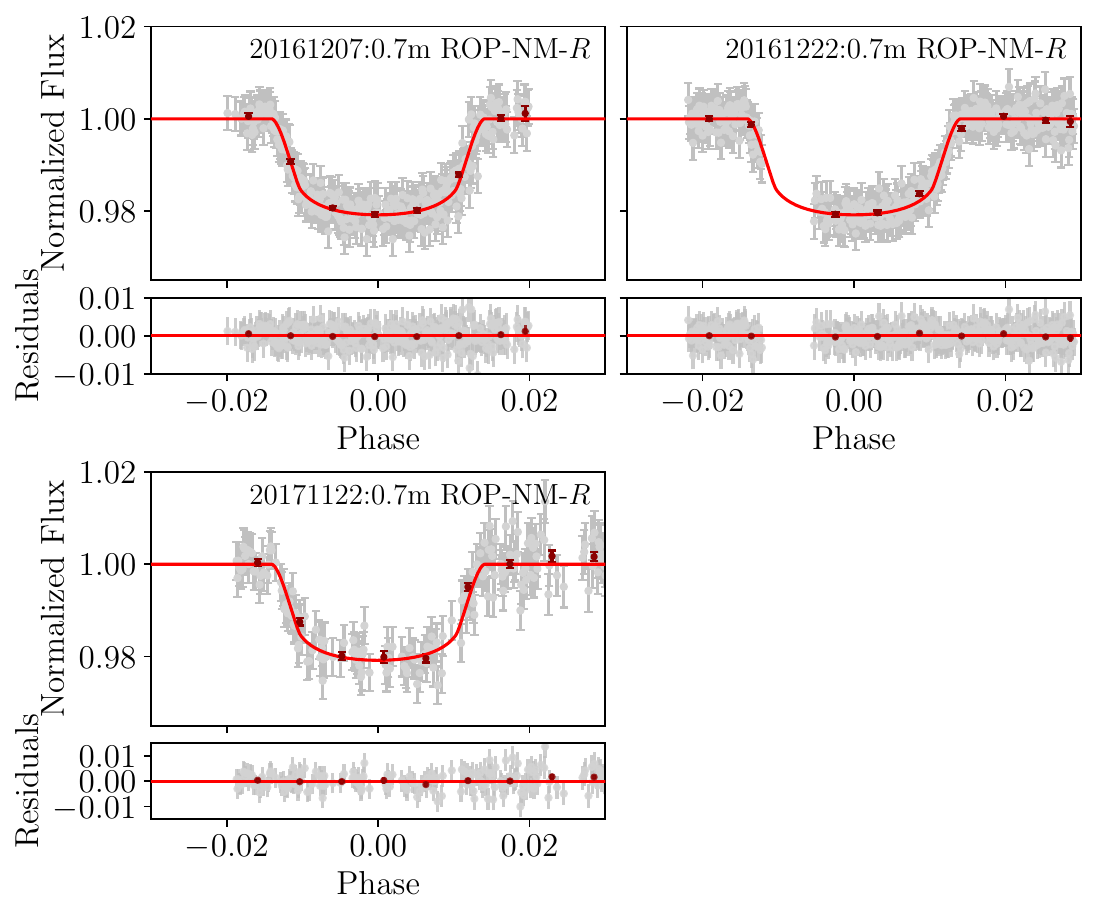} \\
    \includegraphics[scale=0.42]{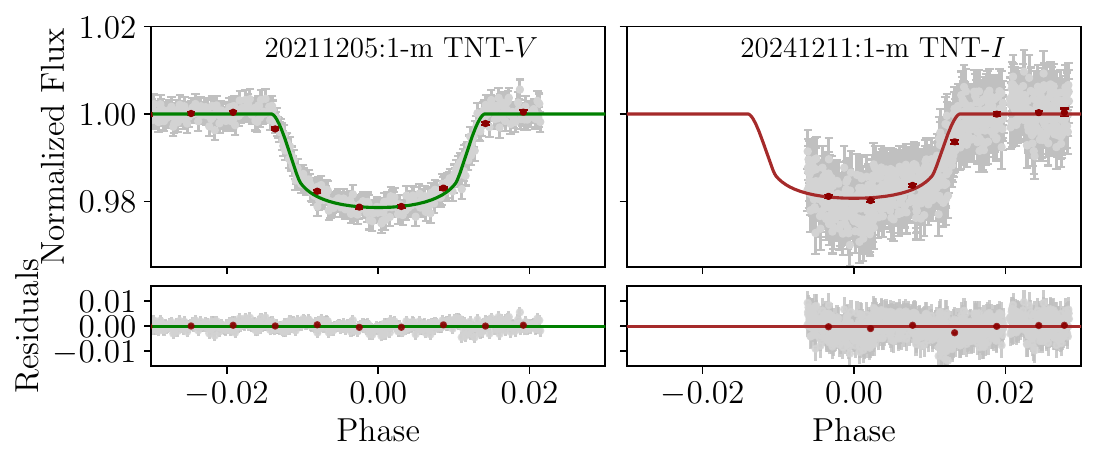}    
\end{tabular}  
    \caption{The normalized, individual transit light curves of WASP-11~b/HAT-P-10~b observed with the SPEARNET telescope network (gray dots, upper panel), including TRT-SRO, ROP-CC, TRT-GAO, ROP-NM, and the 1-m TNT. The best-fitting model from \texttt{TransitFit} is shown as a solid line. The corresponding residuals are displayed in the lower panel.}
\label{fig:LCs_TRT}
\end{figure*}

The individual transit light curves of WASP-11~b/HAT-P-10~b observed with the SPEARNET telescope network are shown in \Cref{fig:LCs_TRT}. The plots include the best-fitting models and their corresponding residuals.

\section{Posterior probability distribution of the MCMC transit timing analyses.}
\begin{figure*}[htb]
\begin{tabular}{cc}
    \includegraphics[scale=0.38]{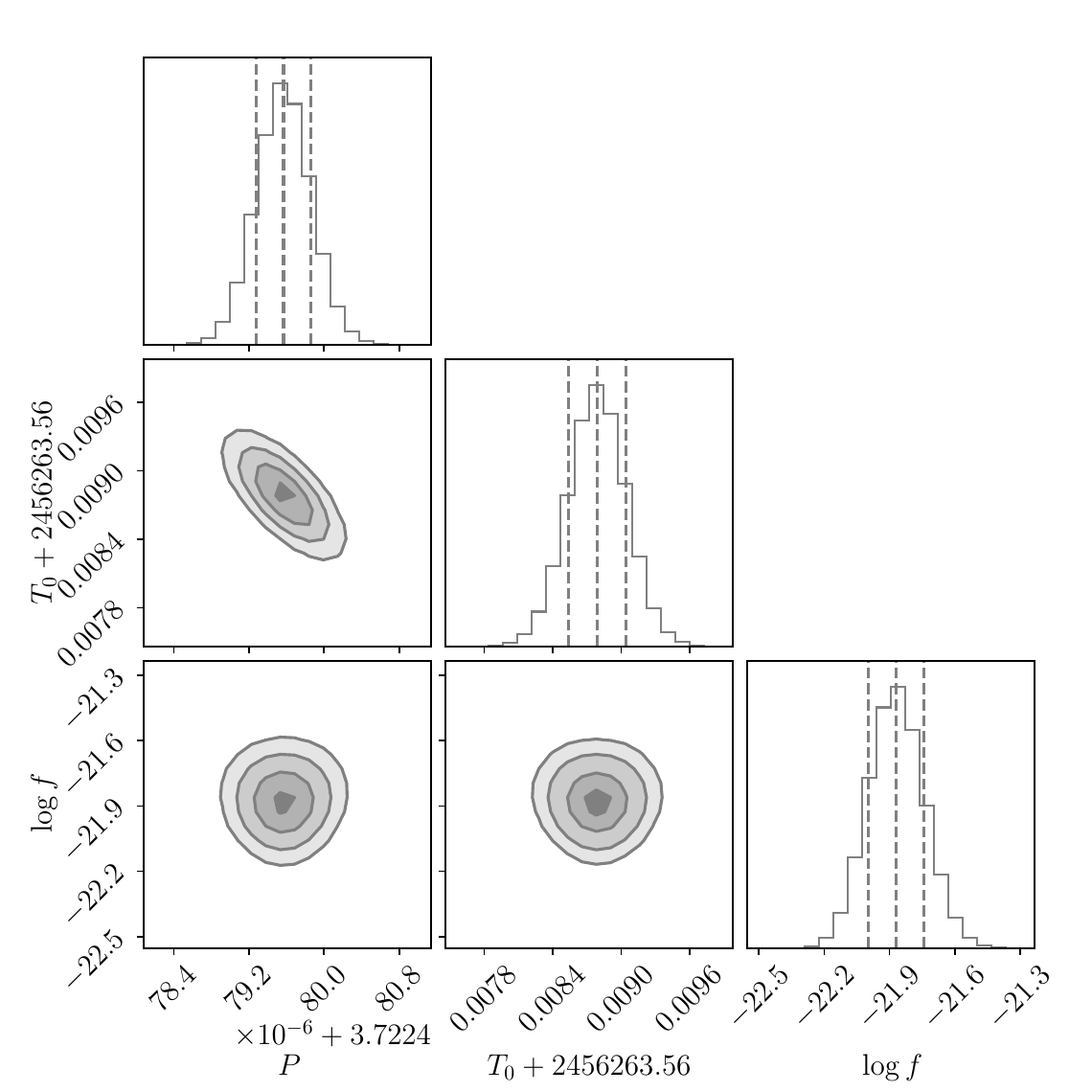} 
    \includegraphics[scale=0.4]{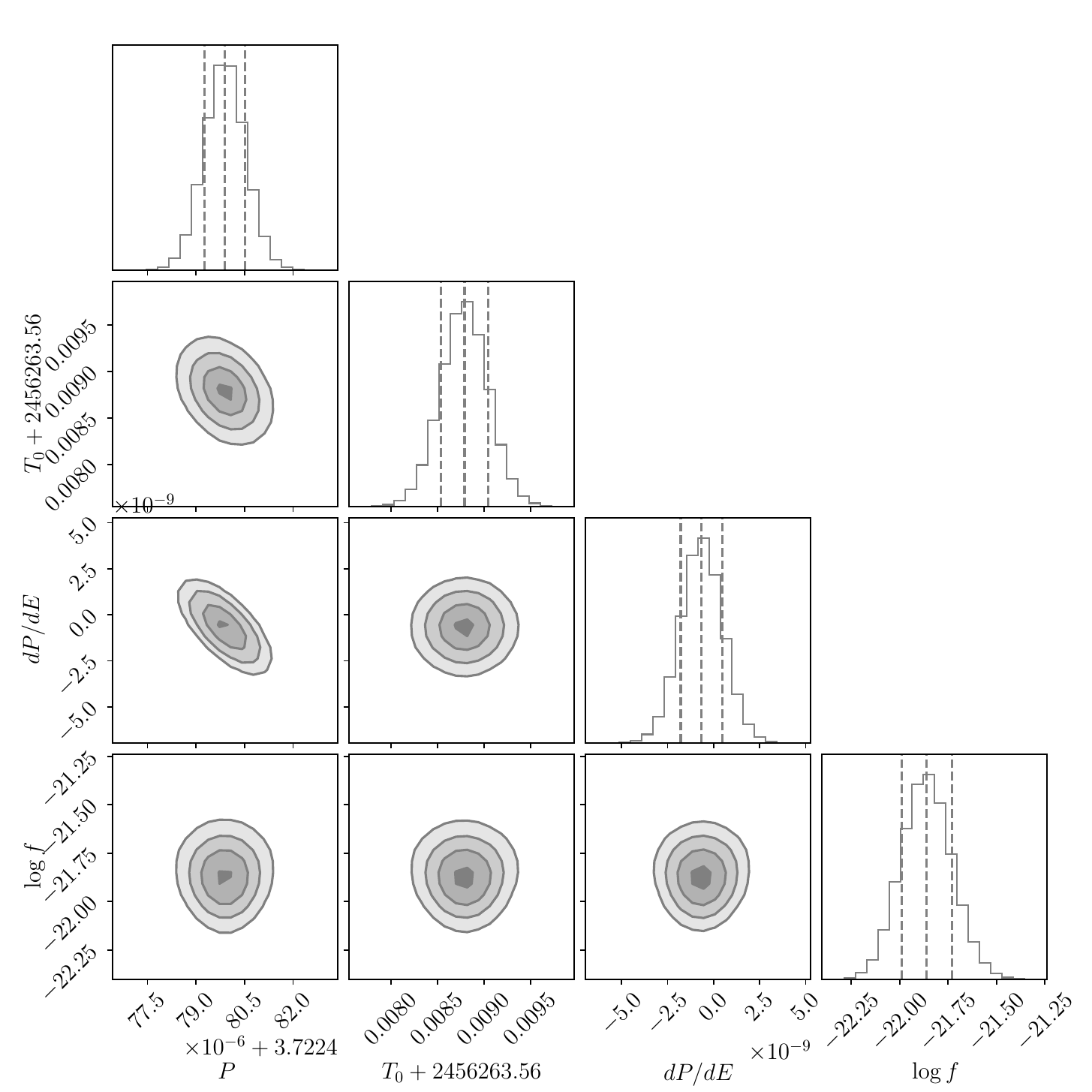} \\ 
    \includegraphics[scale=0.35]{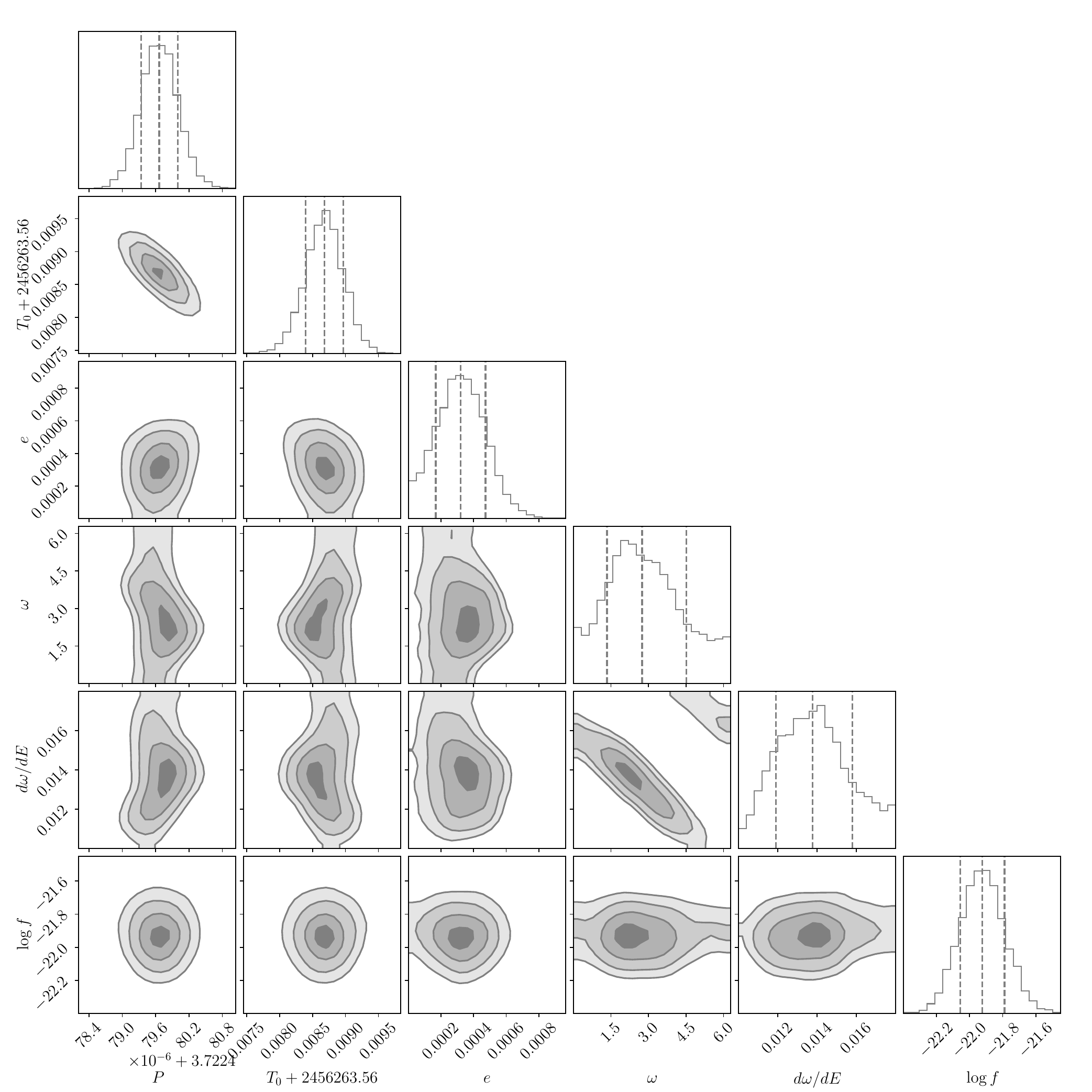}
\end{tabular}
\caption{Posterior probability distribution of the MCMC fitting parameters for the constant-period, orbital decay and apsidal precession models, respectively.}
\label{fig:liDe_mcmc}
\end{figure*}

This section presents the corner plots illustrating the posterior probability distributions derived from the MCMC transit timing analyses for the three models, as shown in \Cref{fig:liDe_mcmc}.

\section{Posterior Probability Distribution from Atmospheric Modeling with \texttt{PLATON}}
In this section, we present the retrieved posterior distributions of the atmospheric parameters for WASP-11~b/HAT-P-10~b. These were obtained using the \texttt{PLATON} code, which employs a nested sampling algorithm for the retrieval process.
\begin{figure*}[htb]
\begin{center}
\includegraphics[scale=0.38]{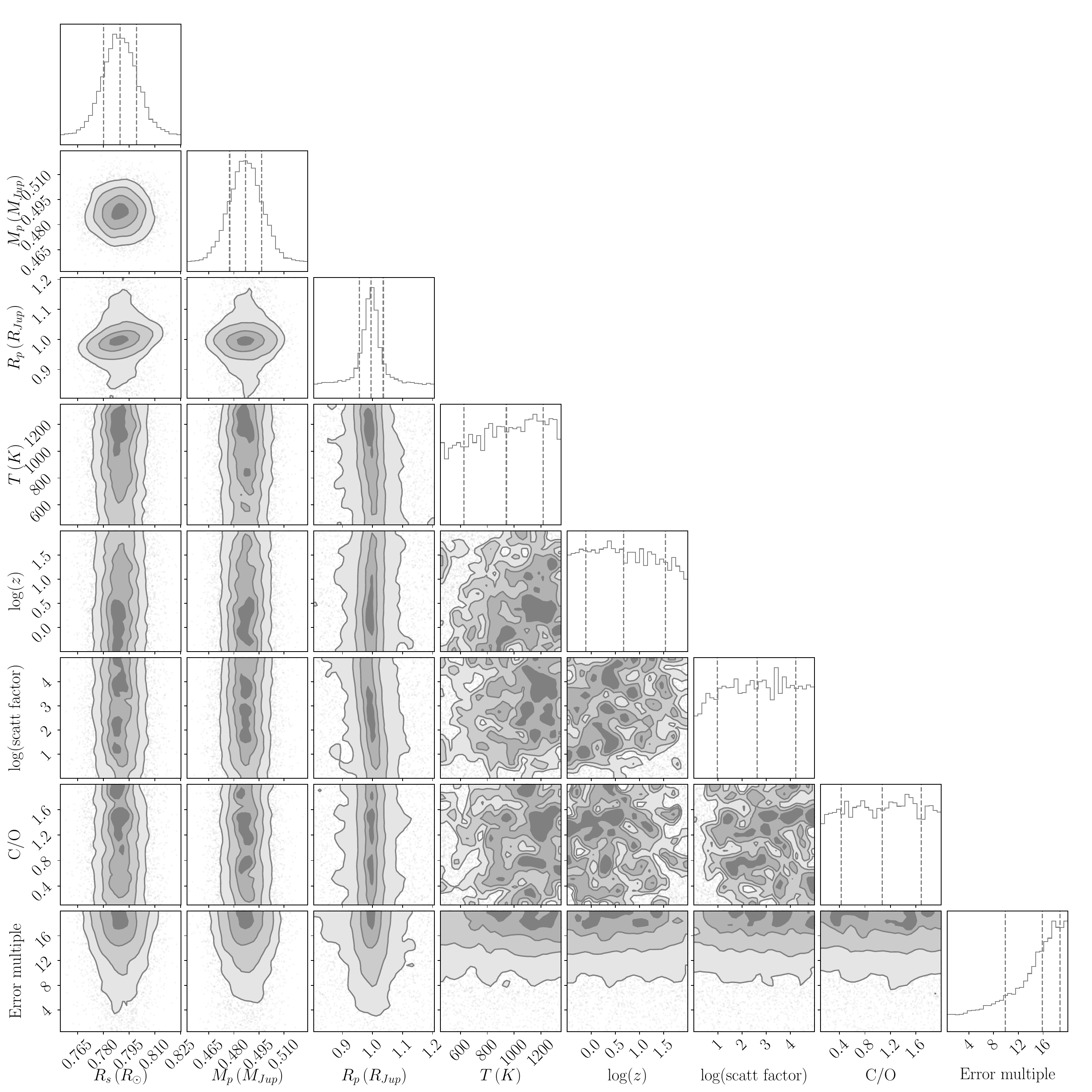} \\    
\caption{The posterior distributions for the atmospheric parameters of WASP-11~b/HAT-P-10~b retrieved using \texttt{PLATON}.}
\label{fig:contour-spec}
\end{center}
\end{figure*}


\bibliography{WASP11b}{}

@string{mnras="{Mon. Not. R. Astr. Soc.}"}

@string{nat="{Nature}"}

@string{apj="{Astrophys. J.}"}

@string{apjl="{Astrophys. J. {Letters}}"}

@string{apjs="{Astrophys. J. Suppl.}"}

@string{pasp="{Publs. Astr. Soc. Pacif.}"}

@string{aas="{Astr. Astrophys. Suppl.}"}

@string{aj="{Astron. J.}"}

@string{actaa="{Acta Astronomica}"}

@string{apss="{Astrophys. Space Sci.}"}

@string{baj="{Berliner Astron. Jahrb.}"}

@ARTICLE{bakos2009,
       author = {{Bakos}, G. {\'A}. and {P{\'a}l}, A. and {Torres}, G. and {Sip{\H{o}}cz}, B. and {Latham}, D.~W. and {Noyes}, R.~W. and {Kov{\'a}cs}, G{\'e}za and {Hartman}, J. and {Esquerdo}, G.~A. and {Fischer}, D.~A. and {Johnson}, J.~A. and {Marcy}, G.~W. and {Butler}, R.~P. and {Howard}, A.~W. and {Sasselov}, D.~D. and {Kov{\'a}cs}, G{\'a}bor and {Stefanik}, R.~P. and {L{\'a}z{\'a}r}, J. and {Papp}, I. and {S{\'a}ri}, P.},
        title = "{HAT-P-10b: A Light and Moderately Hot Jupiter Transiting A K Dwarf}",
      journal = {\apj},
         year = 2009,
        month = may,
       volume = {696},
       number = {2},
        pages = {1950-1955},
          doi = {10.1088/0004-637X/696/2/1950},
archivePrefix = {arXiv},
       eprint = {0809.4295},
 primaryClass = {astro-ph},
       adsurl = {https://ui.adsabs.harvard.edu/abs/2009ApJ...696.1950B}
}

@ARTICLE{jiang2021,
       author = {{Jiang}, C. and {Chen}, G. and {Pall{\'e}}, E. and {Murgas}, F. and {Parviainen}, H. and {Yan}, F. and {Ma}, Y.},
        title = "{Evidence for stellar contamination in the transmission spectra of HAT-P-12b}",
      journal = {\aap},
         year = 2021,
        month = dec,
       volume = {656},
          eid = {A114},
        pages = {A114},
          doi = {10.1051/0004-6361/202141824},
archivePrefix = {arXiv},
       eprint = {2109.11235},
 primaryClass = {astro-ph.EP},
       adsurl = {https://ui.adsabs.harvard.edu/abs/2021A&A...656A.114J}
}

@ARTICLE{nikolov2015,
       author = {{Nikolov}, N. and {Sing}, D.~K. and {Burrows}, A.~S. and {Fortney}, J.~J. and {Henry}, G.~W. and {Pont}, F. and {Ballester}, G.~E. and {Aigrain}, S. and {Wilson}, P.~A. and {Huitson}, C.~M. and {Gibson}, N.~P. and {D{\'e}sert}, J. -M. and {Lecavelier Des Etangs}, A. and {Showman}, A.~P. and {Vidal-Madjar}, A. and {Wakeford}, H.~R. and {Zahnle}, K.},
        title = "{HST hot-Jupiter transmission spectral survey: haze in the atmosphere of WASP-6b}",
      journal = {\mnras},
         year = 2015,
        month = feb,
       volume = {447},
       number = {1},
        pages = {463-478},
          doi = {10.1093/mnras/stu2433},
archivePrefix = {arXiv},
       eprint = {1411.4567},
 primaryClass = {astro-ph.SR},
       adsurl = {https://ui.adsabs.harvard.edu/abs/2015MNRAS.447..463N}
}

@ARTICLE{carter2020,
       author = {{Carter}, Aarynn L. and {Nikolov}, Nikolay and {Sing}, David K. and {Alam}, Munazza K. and {Goyal}, Jayesh M. and {Mikal-Evans}, Thomas and {Wakeford}, Hannah R. and {Henry}, Gregory W. and {Morrell}, Sam and {L{\'o}pez-Morales}, Mercedes and {Smalley}, Barry and {Lavvas}, Panayotis and {Barstow}, Joanna K. and {Garc{\'\i}a Mu{\~n}oz}, Antonio and {Gibson}, Neale P. and {Wilson}, Paul A.},
        title = "{Detection of Na, K, and H$_{2}$O in the hazy atmosphere of WASP-6b}",
      journal = {\mnras},
         year = 2020,
        month = jun,
       volume = {494},
       number = {4},
        pages = {5449-5472},
          doi = {10.1093/mnras/staa1078},
archivePrefix = {arXiv},
       eprint = {1911.12628},
 primaryClass = {astro-ph.EP},
       adsurl = {https://ui.adsabs.harvard.edu/abs/2020MNRAS.494.5449C}
}

@ARTICLE{grubel2025,
       author = {{Gr{\"u}bel}, Fabian and {Molaverdikhani}, Karan and {Ercolano}, Barbara and {Rab}, Christian and {Trapp}, Oliver and {Dubey}, Dwaipayan and {Arenales-Lope}, Rosa},
        title = "{Detectability of polycyclic aromatic hydrocarbons in the atmosphere of WASP-6 b with JWST NIRSpec PRISM}",
      journal = {\mnras},
         year = 2025,
        month = jan,
       volume = {536},
       number = {1},
        pages = {324-339},
          doi = {10.1093/mnras/stae2532},
archivePrefix = {arXiv},
       eprint = {2411.07861},
 primaryClass = {astro-ph.EP},
       adsurl = {https://ui.adsabs.harvard.edu/abs/2025MNRAS.536..324G}
}

@ARTICLE{mancini2015,
       author = {{Mancini}, L. and {Esposito}, M. and {Covino}, E. and {Raia}, G. and {Southworth}, J. and {Tregloan-Reed}, J. and {Biazzo}, K. and {Bonomo}, A.~S. and {Desidera}, S. and {Lanza}, A.~F. and {Maciejewski}, G. and {Poretti}, E. and {Sozzetti}, A. and {Borsa}, F. and {Bruni}, I. and {Ciceri}, S. and {Claudi}, R. and {Cosentino}, R. and {Gratton}, R. and {Martinez Fiorenzano}, A.~F. and {Lodato}, G. and {Lorenzi}, V. and {Marzari}, F. and {Murabito}, S. and {Affer}, L. and {Bignamini}, A. and {Bedin}, L.~R. and {Boccato}, C. and {Damasso}, M. and {Henning}, Th. and {Maggio}, A. and {Micela}, G. and {Molinari}, E. and {Pagano}, I. and {Piotto}, G. and {Rainer}, M. and {Scandariato}, G. and {Smareglia}, R. and {Zanmar Sanchez}, R.},
        title = "{The GAPS Programme with HARPS-N at TNG. VIII. Observations of the Rossiter-McLaughlin effect and characterisation of the transiting planetary systems HAT-P-36 and WASP-11/HAT-P-10}",
      journal = {\aap},
         year = 2015,
        month = jul,
       volume = {579},
          eid = {A136},
        pages = {A136},
          doi = {10.1051/0004-6361/201526030},
archivePrefix = {arXiv},
       eprint = {1503.01787},
 primaryClass = {astro-ph.EP},
       adsurl = {https://ui.adsabs.harvard.edu/abs/2015A&A...579A.136M}
}

@ARTICLE{maciejewski2023,
       author = {{Maciejewski}, G. and {Fern{\'a}ndez}, M. and {Sota}, A. and {Amado}, P.~J. and {Ohlert}, J. and {Bischoff}, R. and {Stenglein}, W. and {Mugrauer}, M. and {Michel}, K. -U. and {Golonka}, J. and {Blanco Solsona}, A. and {Lape{\~n}a}, E. and {Molins Freire}, J. and {de los R{\'\i}os Curieses}, A. and {Temprano Sicilia}, J.~A.},
        title = "{Search for Planets in Hot Jupiter Systems with Multi-Sector TESS Photometry. III. A Study of Ten Systems Enhanced with New Ground-Based Photometry}",
      journal = {\actaa},
         year = 2023,
        month = jul,
       volume = {73},
       number = {1},
        pages = {57-86},
          doi = {10.32023/0001-5237/73.1.4},
archivePrefix = {arXiv},
       eprint = {2307.00538},
 primaryClass = {astro-ph.EP},
       adsurl = {https://ui.adsabs.harvard.edu/abs/2023AcA....73...57M}
}

@INPROCEEDINGS{jenkins2016,
       author = {{Jenkins}, Jon M. and {Twicken}, Joseph D. and {McCauliff}, Sean and {Campbell}, Jennifer and {Sanderfer}, Dwight and {Lung}, David and {Mansouri-Samani}, Masoud and {Girouard}, Forrest and {Tenenbaum}, Peter and {Klaus}, Todd and {Smith}, Jeffrey C. and {Caldwell}, Douglas A. and {Chacon}, A.~D. and {Henze}, Christopher and {Heiges}, Cory and {Latham}, David W. and {Morgan}, Edward and {Swade}, Daryl and {Rinehart}, Stephen and {Vanderspek}, Roland},
        title = "{The TESS science processing operations center}",
    booktitle = {Software and Cyberinfrastructure for Astronomy IV},
         year = 2016,
       editor = {{Chiozzi}, Gianluca and {Guzman}, Juan C.},
       series = {Society of Photo-Optical Instrumentation Engineers (SPIE) Conference Series},
       volume = {9913},
        month = aug,
          eid = {99133E},
        pages = {99133E},
          doi = {10.1117/12.2233418},
       adsurl = {https://ui.adsabs.harvard.edu/abs/2016SPIE.9913E..3EJ}
}

@ARTICLE{hayes2024,
       author = {{Hayes}, J.~J.~C. and {Priyadarshi}, A. and {Kerins}, E. and {Awiphan}, S. and {McDonald}, I. and {A-Thano}, N. and {Morgan}, J.~S. and {Humpage}, A. and {Charles}, S. and {Wright}, M. and {Joshi}, Y.~C. and {Jiang}, Ing-Guey and {Inyanya}, T. and {Padjaroen}, T. and {Munsaket}, P. and {Chuanraksasat}, P. and {Komonjinda}, S. and {Kittara}, P. and {Dhillon}, V.~S. and {Marsh}, T.~R. and {Reichart}, D.~E. and {Poshyachinda}, S. and {SPEARNET Collaboration}},
        title = "{TransitFit: combined multi-instrument exoplanet transit fitting for JWST, HST, and ground-based transmission spectroscopy studies}",
      journal = {\mnras},
         year = 2024,
        month = jan,
       volume = {527},
       number = {3},
        pages = {4936-4954},
          doi = {10.1093/mnras/stad3353},
archivePrefix = {arXiv},
       eprint = {2103.12139},
 primaryClass = {astro-ph.EP},
       adsurl = {https://ui.adsabs.harvard.edu/abs/2024MNRAS.527.4936H}
}

@ARTICLE{athano2023,
       author = {{A-thano}, Napaporn and {Awiphan}, Supachai and {Jiang}, Ing-Guey and {Kerins}, Eamonn and {Priyadarshi}, Akshay and {McDonald}, Iain and {Joshi}, Yogesh C. and {Chulikorn}, Thansuda and {Hayes}, Joshua J.~C. and {Charles}, Stephen and {Huang}, Chung-Kai and {Rattanamala}, Ronnakrit and {Yeh}, Li-Chin and {Dhillon}, Vik S.},
        title = "{Revisiting the Transit Timing and Atmosphere Characterization of the Neptune-mass Planet HAT-P-26 b}",
      journal = {\aj},
         year = 2023,
        month = dec,
       volume = {166},
       number = {6},
          eid = {223},
        pages = {223},
          doi = {10.3847/1538-3881/acfeea},
archivePrefix = {arXiv},
       eprint = {2303.03610},
 primaryClass = {astro-ph.EP},
       adsurl = {https://ui.adsabs.harvard.edu/abs/2023AJ....166..223A}
}

@ARTICLE{dhillon2014,
       author = {{Dhillon}, V.~S. and {Marsh}, T.~R. and {Atkinson}, D.~C. and {Bezawada}, N. and {Bours}, M.~C.~P. and {Copperwheat}, C.~M. and {Gamble}, T. and {Hardy}, L.~K. and {Hickman}, R.~D.~H. and {Irawati}, P. and {Ives}, D.~J. and {Kerry}, P. and {Leckngam}, A. and {Littlefair}, S.~P. and {McLay}, S.~A. and {O'Brien}, K. and {Peacocke}, P.~T. and {Poshyachinda}, S. and {Richichi}, A. and {Soonthornthum}, B. and {Vick}, A.},
        title = "{ULTRASPEC: a high-speed imaging photometer on the 2.4-m Thai National Telescope}",
      journal = {\mnras},
         year = 2014,
        month = nov,
       volume = {444},
       number = {4},
        pages = {4009-4021},
          doi = {10.1093/mnras/stu1660},
archivePrefix = {arXiv},
       eprint = {1408.2733},
 primaryClass = {astro-ph.IM},
       adsurl = {https://ui.adsabs.harvard.edu/abs/2014MNRAS.444.4009D}
}

@ARTICLE{lang2010,
       author = {{Lang}, Dustin and {Hogg}, David W. and {Mierle}, Keir and {Blanton}, Michael and {Roweis}, Sam},
        title = "{Astrometry.net: Blind Astrometric Calibration of Arbitrary Astronomical Images}",
      journal = {\aj},
         year = 2010,
        month = may,
       volume = {139},
       number = {5},
        pages = {1782-1800},
          doi = {10.1088/0004-6256/139/5/1782},
archivePrefix = {arXiv},
       eprint = {0910.2233},
 primaryClass = {astro-ph.IM},
       adsurl = {https://ui.adsabs.harvard.edu/abs/2010AJ....139.1782L}
}

@article{Kano2018,
	doi = {10.3847/2515-5172/aaa4b7},
	url = {https://doi.org/10.3847/2515-5172/aaa4b7},
	year = 2018,
	month = {jan},
	publisher = {American Astronomical Society},
	volume = {2},
	number = {1},
	pages = {4},
	author = {Shubham Kanodia and Jason Wright},
	title = {Python Leap Second Management and Implementation of Precise Barycentric Correction (barycorrpy)},
	journal = {Research Notes of the {AAS}}
}

@ARTICLE{bertin1996,
       author = {{Bertin}, E. and {Arnouts}, S.},
        title = "{SExtractor: Software for source extraction.}",
      journal = {\aaps},
         year = 1996,
        month = jun,
       volume = {117},
        pages = {393-404},
          doi = {10.1051/aas:1996164},
       adsurl = {https://ui.adsabs.harvard.edu/abs/1996A&AS..117..393B}
}

@INPROCEEDINGS{tody1986,
       author = {{Tody}, Doug},
        title = "{The IRAF Data Reduction and Analysis System}",
    booktitle = {Instrumentation in astronomy VI},
         year = 1986,
       editor = {{Crawford}, David L.},
       series = {Society of Photo-Optical Instrumentation Engineers (SPIE) Conference Series},
       volume = {627},
        month = jan,
        pages = {733},
          doi = {10.1117/12.968154},
       adsurl = {https://ui.adsabs.harvard.edu/abs/1986SPIE..627..733T}
}

@INPROCEEDINGS{tody1993,
       author = {{Tody}, Doug},
        title = "{IRAF in the Nineties}",
    booktitle = {Astronomical Data Analysis Software and Systems II},
         year = 1993,
       editor = {{Hanisch}, R.~J. and {Brissenden}, R.~J.~V. and {Barnes}, J.},
       series = {Astronomical Society of the Pacific Conference Series},
       volume = {52},
        month = jan,
        pages = {173},
       adsurl = {https://ui.adsabs.harvard.edu/abs/1993ASPC...52..173T}
}

@article{Fulton2011,
	doi = {10.1088/0004-6256/142/3/84},
	url = {https://doi.org/10.1088/0004-6256/142/3/84},
	year = 2011,
	month = {aug},
	publisher = {American Astronomical Society},
	volume = {142},
	number = {3},
	pages = {84},
	author = {Benjamin J. Fulton and Avi Shporer and Joshua N. Winn and Matthew J. Holman and Andr{\'{a}}s P{\'{a}}l and J. Zachary Gazak},
	title = {{LONG}-{TERM} {TRANSIT} {TIMING} {MONITORING} {AND} {REFINED} {LIGHT} {CURVE} {PARAMETERS} {OF} {HAT}-P-13b},
	journal = {The Astronomical Journal}
}

@ARTICLE{kreidberg2015,
       author = {{Kreidberg}, Laura},
        title = "{batman: BAsic Transit Model cAlculatioN in Python}",
      journal = {\pasp},
         year = 2015,
        month = nov,
       volume = {127},
       number = {957},
        pages = {1161},
          doi = {10.1086/683602},
archivePrefix = {arXiv},
       eprint = {1507.08285},
 primaryClass = {astro-ph.EP},
       adsurl = {https://ui.adsabs.harvard.edu/abs/2015PASP..127.1161K}
}

@ARTICLE{speagle2020,
       author = {{Speagle}, Joshua S.},
        title = "{DYNESTY: a dynamic nested sampling package for estimating Bayesian posteriors and evidences}",
      journal = {\mnras},
         year = 2020,
        month = apr,
       volume = {493},
       number = {3},
        pages = {3132-3158},
          doi = {10.1093/mnras/staa278},
archivePrefix = {arXiv},
       eprint = {1904.02180},
 primaryClass = {astro-ph.IM},
       adsurl = {https://ui.adsabs.harvard.edu/abs/2020MNRAS.493.3132S}
}

@ARTICLE{husser2013,
       author = {{Husser}, T. -O. and {Wende-von Berg}, S. and {Dreizler}, S. and {Homeier}, D. and {Reiners}, A. and {Barman}, T. and {Hauschildt}, P.~H.},
        title = "{A new extensive library of PHOENIX stellar atmospheres and synthetic spectra}",
      journal = {\aap},
         year = 2013,
        month = may,
       volume = {553},
          eid = {A6},
        pages = {A6},
          doi = {10.1051/0004-6361/201219058},
archivePrefix = {arXiv},
       eprint = {1303.5632},
 primaryClass = {astro-ph.SR},
       adsurl = {https://ui.adsabs.harvard.edu/abs/2013A&A...553A...6H}
}

@ARTICLE{parviainen2015,
       author = {{Parviainen}, H. and {Aigrain}, S.},
        title = "{LDTK: Limb Darkening Toolkit}",
      journal = {\mnras},
         year = 2015,
        month = nov,
       volume = {453},
       number = {4},
        pages = {3821-3826},
          doi = {10.1093/mnras/stv1857},
archivePrefix = {arXiv},
       eprint = {1508.02634},
 primaryClass = {astro-ph.EP},
       adsurl = {https://ui.adsabs.harvard.edu/abs/2015MNRAS.453.3821P}
}

@ARTICLE{bonomo2017,
       author = {{Bonomo}, A.~S. and {Desidera}, S. and {Benatti}, S. and {Borsa}, F. and {Crespi}, S. and {Damasso}, M. and {Lanza}, A.~F. and {Sozzetti}, A. and {Lodato}, G. and {Marzari}, F. and {Boccato}, C. and {Claudi}, R.~U. and {Cosentino}, R. and {Covino}, E. and {Gratton}, R. and {Maggio}, A. and {Micela}, G. and {Molinari}, E. and {Pagano}, I. and {Piotto}, G. and {Poretti}, E. and {Smareglia}, R. and {Affer}, L. and {Biazzo}, K. and {Bignamini}, A. and {Esposito}, M. and {Giacobbe}, P. and {H{\'e}brard}, G. and {Malavolta}, L. and {Maldonado}, J. and {Mancini}, L. and {Martinez Fiorenzano}, A. and {Masiero}, S. and {Nascimbeni}, V. and {Pedani}, M. and {Rainer}, M. and {Scandariato}, G.},
        title = "{The GAPS Programme with HARPS-N at TNG . XIV. Investigating giant planet migration history via improved eccentricity and mass determination for 231 transiting planets}",
      journal = {\aap},
         year = 2017,
        month = jun,
       volume = {602},
          eid = {A107},
        pages = {A107},
          doi = {10.1051/0004-6361/201629882},
archivePrefix = {arXiv},
       eprint = {1704.00373},
 primaryClass = {astro-ph.EP},
       adsurl = {https://ui.adsabs.harvard.edu/abs/2017A&A...602A.107B}
}

@ARTICLE{wang2014,
       author = {{Wang}, Xiao-bin and {Gu}, Sheng-hong and {Collier Cameron}, Andrew and {Wang}, Yi-bo and {Hui}, Ho-Keung and {Kwok}, Chi-Tai and {Yeung}, Bill and {Leung}, Kam-Cheung},
        title = "{The Refined Physical Properties of Transiting Exoplanetary System WASP-11/HAT-P-10}",
      journal = {\aj},
         year = 2014,
        month = apr,
       volume = {147},
       number = {4},
          eid = {92},
        pages = {92},
          doi = {10.1088/0004-6256/147/4/92},
       adsurl = {https://ui.adsabs.harvard.edu/abs/2014AJ....147...92W}
}

@ARTICLE{wang2024,
       author = {{Wang}, Wenqin and {Zhang}, Zixin and {Chen}, Zhangliang and {Wang}, Yonghao and {Yu}, Cong and {Ma}, Bo},
        title = "{Long-term Variations in the Orbital Period of Hot Jupiters from Transit-timing Analysis Using TESS Survey Data}",
      journal = {\apjs},
         year = 2024,
        month = jan,
       volume = {270},
       number = {1},
          eid = {14},
        pages = {14},
          doi = {10.3847/1538-4365/ad0847},
archivePrefix = {arXiv},
       eprint = {2310.17225},
 primaryClass = {astro-ph.EP},
       adsurl = {https://ui.adsabs.harvard.edu/abs/2024ApJS..270...14W}
}

@ARTICLE{yalcin2024,
       author = {{Yal{\c{c}}{\i}nkaya}, S. and {Esmer}, E.~M. and {Ba{\c{s}}t{\"u}rk}, {\"O}. and {Muhaymin}, A. and {Kutluay}, A.~C. and {Silistre}, D. {\.I}. and {Akar}, F. and {Southworth}, J. and {Mancini}, L. and {Davoudi}, F. and {Karamanl{\i}}, E. and {Tezcan}, F. and {Demir}, E. and {Y{\i}lmaz}, D. and {G{\"u}lero{\u{g}}lu}, E. and {Tekin}, M. and {Ta{\c{s}}k{\i}n}, {\.I}. and {Alada{\u{g}}}, Y. and {Sertkan}, E. and {Kurt}, U.~Y. and {Fi{\c{s}}ek}, S. and {Kaptan}, S. and {Ali{\c{s}}}, S. and {Aksaker}, N. and {Yelkenci}, F.~K. and {Tezcan}, C.~T. and {Kaya}, A. and {O{\u{g}}lakkaya}, D. and {Ayd{\i}n}, Z.~S. and {Ye{\c{s}}ilyaprak}, C.},
        title = "{Looking for timing variations in the transits of 16 exoplanets}",
      journal = {\mnras},
         year = 2024,
        month = may,
       volume = {530},
       number = {3},
        pages = {2475-2495},
          doi = {10.1093/mnras/stae854},
archivePrefix = {arXiv},
       eprint = {2403.17690},
 primaryClass = {astro-ph.EP},
       adsurl = {https://ui.adsabs.harvard.edu/abs/2024MNRAS.530.2475Y}
}

@ARTICLE{Er2024,
       author = {{Er}, Huseyin and {Karaman}, Nazl{\i} and {{\"O}zd{\"o}nmez}, Aykut and {Nasiroglu}, {\.I}lham and {G{\"u}rbulak}, B. Batuhan},
        title = "{Investigation on transit observations of the WASP-10 and WASP-11 systems}",
      journal = {\na},
         year = 2024,
        month = apr,
       volume = {107},
          eid = {102138},
        pages = {102138},
          doi = {10.1016/j.newast.2023.102138},
       adsurl = {https://ui.adsabs.harvard.edu/abs/2024NewA..10702138E}
}

@ARTICLE{ivshina2022,
       author = {{Ivshina}, Ekaterina S. and {Winn}, Joshua N.},
        title = "{TESS Transit Timing of Hundreds of Hot Jupiters}",
      journal = {\apjs},
         year = 2022,
        month = apr,
       volume = {259},
       number = {2},
          eid = {62},
        pages = {62},
          doi = {10.3847/1538-4365/ac545b},
archivePrefix = {arXiv},
       eprint = {2202.03401},
 primaryClass = {astro-ph.EP},
       adsurl = {https://ui.adsabs.harvard.edu/abs/2022ApJS..259...62I}
}

@ARTICLE{west2009,
       author = {{West}, R.~G. and {Collier Cameron}, A. and {Hebb}, L. and {Joshi}, Y.~C. and {Pollacco}, D. and {Simpson}, E. and {Skillen}, I. and {Stempels}, H.~C. and {Wheatley}, P.~J. and {Wilson}, D. and {Anderson}, D. and {Bentley}, S. and {Bouchy}, F. and {Christian}, D. and {Enoch}, B. and {Gibson}, N. and {H{\'e}brard}, G. and {Hellier}, C. and {Loeillet}, B. and {Mayor}, M. and {Maxted}, P. and {McDonald}, I. and {Moutou}, C. and {Pont}, F. and {Queloz}, D. and {Smith}, A.~M.~S. and {Smalley}, B. and {Street}, R.~A. and {Udry}, S.},
        title = "{The sub-Jupiter mass transiting exoplanet WASP-11b}",
      journal = {\aap},
         year = 2009,
        month = jul,
       volume = {502},
       number = {1},
        pages = {395-400},
          doi = {10.1051/0004-6361/200810973},
archivePrefix = {arXiv},
       eprint = {0809.4597},
 primaryClass = {astro-ph},
       adsurl = {https://ui.adsabs.harvard.edu/abs/2009A&A...502..395W}
}

@ARTICLE{zech2009,
       author = {{Zechmeister}, M. and {K{\"u}rster}, M.},
        title = "{The generalised Lomb-Scargle periodogram. A new formalism for the floating-mean and Keplerian periodograms}",
      journal = {\aap},
         year = 2009,
        month = mar,
       volume = {496},
       number = {2},
        pages = {577-584},
          doi = {10.1051/0004-6361:200811296},
archivePrefix = {arXiv},
       eprint = {0901.2573},
 primaryClass = {astro-ph.IM},
       adsurl = {https://ui.adsabs.harvard.edu/abs/2009A&A...496..577Z}
}

@MISC{pya2019,
       author = {{Czesla}, Stefan and {Schr{\"o}ter}, Sebastian and
         {Schneider}, Christian P. and {Huber}, Klaus F. and {Pfeifer}, Fabian and
         {Andreasen}, Daniel T. and {Zechmeister}, Mathias},
        title = "{PyA: Python astronomy-related packages}",
         year = "2019",
        month = "Jun",
          eid = {ascl:1906.010},
        pages = {ascl:1906.010},
archivePrefix = {ascl},
       eprint = {1906.010},
       adsurl = {https://ui.adsabs.harvard.edu/abs/2019ascl.soft06010C}
}

@ARTICLE{kokori2023,
       author = {{Kokori}, A. and {Tsiaras}, A. and {Edwards}, B. and {Jones}, A. and {Pantelidou}, G. and {Tinetti}, G. and {Bewersdorff}, L. and {Iliadou}, A. and {Jongen}, Y. and {Lekkas}, G. and {Nastasi}, A. and {Poultourtzidis}, E. and {Sidiropoulos}, C. and {Walter}, F. and {W{\"u}nsche}, A. and {Abraham}, R. and {Agnihotri}, V.~K. and {Albanesi}, R. and {Arce-Mansego}, E. and {Arnot}, D. and {Audejean}, M. and {Aumasson}, C. and {Bachschmidt}, M. and {Baj}, G. and {Barroy}, P.~R. and {Belinski}, A.~A. and {Bennett}, D. and {Benni}, P. and {Bernacki}, K. and {Betti}, L. and {Biagini}, A. and {Bosch}, P. and {Brandebourg}, P. and {Br{\'a}t}, L. and {Bretton}, M. and {Brincat}, S.~M. and {Brouillard}, S. and {Bruzas}, A. and {Bruzzone}, A. and {Buckland}, R.~A. and {Cal{\'o}}, M. and {Campos}, F. and {Carre{\~n}o}, A. and {Carrion Rodrigo}, J.~A. and {Casali}, R. and {Casalnuovo}, G. and {Cataneo}, M. and {Chang}, C. -M. and {Changeat}, L. and {Chowdhury}, V. and {Ciantini}, R. and {Cilluffo}, M. and {Coliac}, J. -F. and {Conzo}, G. and {Correa}, M. and {Coulon}, G. and {Crouzet}, N. and {Crow}, M.~V. and {Curtis}, I.~A. and {Daniel}, D. and {Dauchet}, B. and {Dawes}, S. and {Deldem}, M. and {Deligeorgopoulos}, D. and {Dransfield}, G. and {Dymock}, R. and {Eenm{\"a}e}, T. and {Esseiva}, N. and {Evans}, P. and {Falco}, C. and {Farf{\'a}n}, R.~G. and {Fern{\'a}ndez-Laj{\'u}s}, E. and {Ferratfiat}, S. and {Ferreira}, S.~L. and {Ferretti}, A. and {Fio{\l}ka}, J. and {Fowler}, M. and {Futcher}, S.~R. and {Gabellini}, D. and {Gainey}, T. and {Gaitan}, J. and {Gajdo{\v{s}}}, P. and {Garc{\'\i}a-S{\'a}nchez}, A. and {Garlitz}, J. and {Gillier}, C. and {Gison}, C. and {Gonzales}, J. and {Gorshanov}, D. and {Grau Horta}, F. and {Grivas}, G. and {Guerra}, P. and {Guillot}, T. and {Haswell}, C.~A. and {Haymes}, T. and {Hentunen}, V. -P. and {Hills}, K. and {Hose}, K. and {Humbert}, T. and {Hurter}, F. and {Hynek}, T. and {Irzyk}, M. and {Jacobsen}, J. and {Jannetta}, A.~L. and {Johnson}, K. and {J{\'o}{\'z}wik-Wabik}, P. and {Kaeouach}, A.~E. and {Kang}, W. and {Kiiskinen}, H. and {Kim}, T. and {Kivila}, {\"U}. and {Koch}, B. and {Kolb}, U. and {Ku{\v{c}}{\'a}kov{\'a}}, H. and {Lai}, S. -P. and {Laloum}, D. and {Lasota}, S. and {Lewis}, L.~A. and {Liakos}, G. -I. and {Libotte}, F. and {Lomoz}, F. and {Lopresti}, C. and {Majewski}, R. and {Malcher}, A. and {Mallonn}, M. and {Mannucci}, M. and {Marchini}, A. and {Mari}, J. -M. and {Marino}, A. and {Marino}, G. and {Mario}, J. -C. and {Marquette}, J. -B. and {Mart{\'\i}nez-Bravo}, F.~A. and {Ma{\v{s}}ek}, M. and {Matassa}, P. and {Michel}, P. and {Michelet}, J. and {Miller}, M. and {Miny}, E. and {Molina}, D. and {Mollier}, T. and {Monteleone}, B. and {Montigiani}, N. and {Morales-Aimar}, M. and {Mortari}, F. and {Morvan}, M. and {Mugnai}, L.~V. and {Murawski}, G. and {Naponiello}, L. and {Naudin}, J. -L. and {Naves}, R. and {N{\'e}el}, D. and {Neito}, R. and {Neveu}, S. and {Noschese}, A. and {{\"O}{\u{g}}men}, Y. and {Ohshima}, O. and {Orbanic}, Z. and {Pace}, E.~P. and {Pantacchini}, C. and {Paschalis}, N.~I. and {Pereira}, C. and {Peretto}, I. and {Perroud}, V. and {Phillips}, M. and {Pintr}, P. and {Pioppa}, J. -B. and {Plazas}, J. and {Poelarends}, A.~J. and {Popowicz}, A. and {Purcell}, J. and {Quinn}, N. and {Raetz}, M. and {Rees}, D. and {Regembal}, F. and {Rocchetto}, M. and {Rocci}, P. -F. and {Rockenbauer}, M. and {Roth}, R. and {Rousselot}, L. and {Rubia}, X. and {Ruocco}, N. and {Russo}, E. and {Salisbury}, M. and {Salvaggio}, F. and {Santos}, A. and {Savage}, J. and {Scaggiante}, F. and {Sedita}, D. and {Shadick}, S. and {Silva}, A.~F. and {Sioulas}, N. and {{\v{S}}koln{\'\i}k}, V. and {Smith}, M. and {Smolka}, M. and {Solmaz}, A. and {Stanbury}, N. and {Stouraitis}, D. and {Tan}, T. -G. and {Theusner}, M. and {Thurston}, G.},
        title = "{ExoClock Project. III. 450 New Exoplanet Ephemerides from Ground and Space Observations}",
      journal = {\apjs},
         year = 2023,
        month = mar,
       volume = {265},
       number = {1},
          eid = {4},
        pages = {4},
          doi = {10.3847/1538-4365/ac9da4},
archivePrefix = {arXiv},
       eprint = {2209.09673},
 primaryClass = {astro-ph.EP},
       adsurl = {https://ui.adsabs.harvard.edu/abs/2023ApJS..265....4K}
}

@ARTICLE{ngo2015,
       author = {{Ngo}, Henry and {Knutson}, Heather A. and {Hinkley}, Sasha and {Crepp}, Justin R. and {Bechter}, Eric B. and {Batygin}, Konstantin and {Howard}, Andrew W. and {Johnson}, John A. and {Morton}, Timothy D. and {Muirhead}, Philip S.},
        title = "{Friends of Hot Jupiters. II. No Correspondence between Hot-jupiter Spin-Orbit Misalignment and the Incidence of Directly Imaged Stellar Companions}",
      journal = {\apj},
         year = 2015,
        month = feb,
       volume = {800},
       number = {2},
          eid = {138},
        pages = {138},
          doi = {10.1088/0004-637X/800/2/138},
archivePrefix = {arXiv},
       eprint = {1501.00013},
 primaryClass = {astro-ph.EP},
       adsurl = {https://ui.adsabs.harvard.edu/abs/2015ApJ...800..138N}
}

@ARTICLE{knu2014ApJ,
       author = {{Knutson}, Heather A. and {Fulton}, Benjamin J. and {Montet}, Benjamin T. and {Kao}, Melodie and {Ngo}, Henry and {Howard}, Andrew W. and {Crepp}, Justin R. and {Hinkley}, Sasha and {Bakos}, Gaspar {\'A}. and {Batygin}, Konstantin and {Johnson}, John Asher and {Morton}, Timothy D. and {Muirhead}, Philip S.},
        title = "{Friends of Hot Jupiters. I. A Radial Velocity Search for Massive, Long-period Companions to Close-in Gas Giant Planets}",
      journal = {\apj},
         year = 2014,
        month = apr,
       volume = {785},
       number = {2},
          eid = {126},
        pages = {126},
          doi = {10.1088/0004-637X/785/2/126},
archivePrefix = {arXiv},
       eprint = {1312.2954},
 primaryClass = {astro-ph.EP},
       adsurl = {https://ui.adsabs.harvard.edu/abs/2014ApJ...785..126K}
}

@INPROCEEDINGS{ricker2014,
       author = {{Ricker}, George R. and {Winn}, Joshua N. and {Vanderspek}, Roland and {Latham}, David W. and {Bakos}, G{\'a}sp{\'a}r. {\'A}. and {Bean}, Jacob L. and {Berta-Thompson}, Zachory K. and {Brown}, Timothy M. and {Buchhave}, Lars and {Butler}, Nathaniel R. and {Butler}, R. Paul and {Chaplin}, William J. and {Charbonneau}, David and {Christensen-Dalsgaard}, J{\o}rgen and {Clampin}, Mark and {Deming}, Drake and {Doty}, John and {De Lee}, Nathan and {Dressing}, Courtney and {Dunham}, E.~W. and {Endl}, Michael and {Fressin}, Francois and {Ge}, Jian and {Henning}, Thomas and {Holman}, Matthew J. and {Howard}, Andrew W. and {Ida}, Shigeru and {Jenkins}, Jon and {Jernigan}, Garrett and {Johnson}, John A. and {Kaltenegger}, Lisa and {Kawai}, Nobuyuki and {Kjeldsen}, Hans and {Laughlin}, Gregory and {Levine}, Alan M. and {Lin}, Douglas and {Lissauer}, Jack J. and {MacQueen}, Phillip and {Marcy}, Geoffrey and {McCullough}, P.~R. and {Morton}, Timothy D. and {Narita}, Norio and {Paegert}, Martin and {Palle}, Enric and {Pepe}, Francesco and {Pepper}, Joshua and {Quirrenbach}, Andreas and {Rinehart}, S.~A. and {Sasselov}, Dimitar and {Sato}, Bun'ei and {Seager}, Sara and {Sozzetti}, Alessandro and {Stassun}, Keivan G. and {Sullivan}, Peter and {Szentgyorgyi}, Andrew and {Torres}, Guillermo and {Udry}, Stephane and {Villasenor}, Joel},
        title = "{Transiting Exoplanet Survey Satellite (TESS)}",
    booktitle = {Space Telescopes and Instrumentation 2014: Optical, Infrared, and Millimeter Wave},
         year = 2014,
       editor = {{Oschmann}, Jacobus M., Jr. and {Clampin}, Mark and {Fazio}, Giovanni G. and {MacEwen}, Howard A.},
       series = {Society of Photo-Optical Instrumentation Engineers (SPIE) Conference Series},
       volume = {9143},
        month = aug,
          eid = {914320},
        pages = {914320},
          doi = {10.1117/12.2063489},
archivePrefix = {arXiv},
       eprint = {1406.0151},
 primaryClass = {astro-ph.EP},
       adsurl = {https://ui.adsabs.harvard.edu/abs/2014SPIE.9143E..20R}
}

@ARTICLE{irwon1952,
       author = {{Irwin}, John B.},
        title = "{The Determination of a Light-Time Orbit.}",
      journal = {\apj},
         year = 1952,
        month = jul,
       volume = {116},
        pages = {211},
          doi = {10.1086/145604},
       adsurl = {https://ui.adsabs.harvard.edu/abs/1952ApJ...116..211I}
}

@ARTICLE{seager2000,
       author = {{Seager}, S. and {Sasselov}, D.~D.},
        title = "{Theoretical Transmission Spectra during Extrasolar Giant Planet Transits}",
      journal = {\apj},
         year = 2000,
        month = jul,
       volume = {537},
       number = {2},
        pages = {916-921},
          doi = {10.1086/309088},
archivePrefix = {arXiv},
       eprint = {astro-ph/9912241},
 primaryClass = {astro-ph},
       adsurl = {https://ui.adsabs.harvard.edu/abs/2000ApJ...537..916S}
}

@ARTICLE{sing2016,
       author = {{Sing}, David K. and {Fortney}, Jonathan J. and {Nikolov}, Nikolay and {Wakeford}, Hannah R. and {Kataria}, Tiffany and {Evans}, Thomas M. and {Aigrain}, Suzanne and {Ballester}, Gilda E. and {Burrows}, Adam S. and {Deming}, Drake and {D{\'e}sert}, Jean-Michel and {Gibson}, Neale P. and {Henry}, Gregory W. and {Huitson}, Catherine M. and {Knutson}, Heather A. and {Lecavelier Des Etangs}, Alain and {Pont}, Frederic and {Showman}, Adam P. and {Vidal-Madjar}, Alfred and {Williamson}, Michael H. and {Wilson}, Paul A.},
        title = "{A continuum from clear to cloudy hot-Jupiter exoplanets without primordial water depletion}",
      journal = {\nat},
         year = 2016,
        month = jan,
       volume = {529},
       number = {7584},
        pages = {59-62},
          doi = {10.1038/nature16068},
archivePrefix = {arXiv},
       eprint = {1512.04341},
 primaryClass = {astro-ph.EP},
       adsurl = {https://ui.adsabs.harvard.edu/abs/2016Natur.529...59S}
}

@ARTICLE{spy2023,
       author = {{Spyratos}, Petros and {Nikolov}, Nikolay K. and {Constantinou}, Savvas and {Southworth}, John and {Madhusudhan}, Nikku and {Sedaghati}, Elyar and {Ehrenreich}, David and {Mancini}, Luigi},
        title = "{A precise blue-optical transmission spectrum from the ground: evidence for haze in the atmosphere of WASP-74b}",
      journal = {\mnras},
         year = 2023,
        month = may,
       volume = {521},
       number = {2},
        pages = {2163-2180},
          doi = {10.1093/mnras/stad637},
archivePrefix = {arXiv},
       eprint = {2302.11495},
 primaryClass = {astro-ph.EP},
       adsurl = {https://ui.adsabs.harvard.edu/abs/2023MNRAS.521.2163S}
}

@ARTICLE{fairman2024,
       author = {{Fairman}, Charlotte and {Wakeford}, Hannah R. and {MacDonald}, Ryan J.},
        title = "{The Importance of Optical Wavelength Data on Atmospheric Retrievals of Exoplanet Transmission Spectra}",
      journal = {\aj},
         year = 2024,
        month = may,
       volume = {167},
       number = {5},
          eid = {240},
        pages = {240},
          doi = {10.3847/1538-3881/ad3454},
archivePrefix = {arXiv},
       eprint = {2403.07801},
 primaryClass = {astro-ph.EP},
       adsurl = {https://ui.adsabs.harvard.edu/abs/2024AJ....167..240F}
}

@ARTICLE{kirk2017,
       author = {{Kirk}, J. and {Wheatley}, P.~J. and {Louden}, T. and {Doyle}, A.~P. and {Skillen}, I. and {McCormac}, J. and {Irwin}, P.~G.~J. and {Karjalainen}, R.},
        title = "{Rayleigh scattering in the transmission spectrum of HAT-P-18b}",
      journal = {\mnras},
         year = 2017,
        month = jul,
       volume = {468},
       number = {4},
        pages = {3907-3916},
          doi = {10.1093/mnras/stx752},
archivePrefix = {arXiv},
       eprint = {1611.06916},
 primaryClass = {astro-ph.EP},
       adsurl = {https://ui.adsabs.harvard.edu/abs/2017MNRAS.468.3907K}
}

@ARTICLE{leca2008,
       author = {{Lecavelier Des Etangs}, A. and {Pont}, F. and {Vidal-Madjar}, A. and {Sing}, D.},
        title = "{Rayleigh scattering in the transit spectrum of HD 189733b}",
      journal = {\aap},
         year = 2008,
        month = apr,
       volume = {481},
       number = {2},
        pages = {L83-L86},
          doi = {10.1051/0004-6361:200809388},
archivePrefix = {arXiv},
       eprint = {0802.3228},
 primaryClass = {astro-ph},
       adsurl = {https://ui.adsabs.harvard.edu/abs/2008A&A...481L..83L}
}

@ARTICLE{lomb1976,
       author = {{Lomb}, N.~R.},
        title = "{Least-Squares Frequency Analysis of Unequally Spaced Data}",
      journal = {\apss},
         year = 1976,
        month = feb,
       volume = {39},
       number = {2},
        pages = {447-462},
          doi = {10.1007/BF00648343},
       adsurl = {https://ui.adsabs.harvard.edu/abs/1976Ap&SS..39..447L}
}

@ARTICLE{zhang2019,
       author = {{Zhang}, Michael and {Chachan}, Yayaati and {Kempton}, Eliza M. -R. and {Knutson}, Heather A.},
        title = "{Forward Modeling and Retrievals with PLATON, a Fast Open-source Tool}",
      journal = {\pasp},
         year = 2019,
        month = mar,
       volume = {131},
       number = {997},
        pages = {034501},
          doi = {10.1088/1538-3873/aaf5ad},
archivePrefix = {arXiv},
       eprint = {1811.11761},
 primaryClass = {astro-ph.EP},
       adsurl = {https://ui.adsabs.harvard.edu/abs/2019PASP..131c4501Z}
}

@ARTICLE{feroz2009,
       author = {{Feroz}, F. and {Hobson}, M.~P. and {Bridges}, M.},
        title = "{MULTINEST: an efficient and robust Bayesian inference tool for cosmology and particle physics}",
      journal = {\mnras},
         year = 2009,
        month = oct,
       volume = {398},
       number = {4},
        pages = {1601-1614},
          doi = {10.1111/j.1365-2966.2009.14548.x},
archivePrefix = {arXiv},
       eprint = {0809.3437},
 primaryClass = {astro-ph},
       adsurl = {https://ui.adsabs.harvard.edu/abs/2009MNRAS.398.1601F}
}

@ARTICLE{gold1966,
       author = {{Goldreich}, Peter and {Soter}, Steven},
        title = "{Q in the Solar System}",
      journal = {\icarus},
         year = 1966,
        month = jan,
       volume = {5},
       number = {1},
        pages = {375-389},
          doi = {10.1016/0019-1035(66)90051-0},
       adsurl = {https://ui.adsabs.harvard.edu/abs/1966Icar....5..375G}
}

@ARTICLE{penev2018,
       author = {{Penev}, Kaloyan and {Bouma}, L.~G. and {Winn}, Joshua N. and {Hartman}, Joel D.},
        title = "{Empirical Tidal Dissipation in Exoplanet Hosts From Tidal Spin-up}",
      journal = {\aj},
         year = 2018,
        month = apr,
       volume = {155},
       number = {4},
          eid = {165},
        pages = {165},
          doi = {10.3847/1538-3881/aaaf71},
archivePrefix = {arXiv},
       eprint = {1802.05269},
 primaryClass = {astro-ph.SR},
       adsurl = {https://ui.adsabs.harvard.edu/abs/2018AJ....155..165P}
}

@ARTICLE{wong2020,
       author = {{Wong}, Ian and {Benneke}, Bj{\"o}rn and {Gao}, Peter and {Knutson}, Heather A. and {Chachan}, Yayaati and {Henry}, Gregory W. and {Deming}, Drake and {Kataria}, Tiffany and {Lee}, Elspeth K.~H. and {Nikolov}, Nikolay and {Sing}, David K. and {Ballester}, Gilda E. and {Baskin}, Nathaniel J. and {Wakeford}, Hannah R. and {Williamson}, Michael H.},
        title = "{Optical to Near-infrared Transmission Spectrum of the Warm Sub-Saturn HAT-P-12b}",
      journal = {\aj},
         year = 2020,
        month = may,
       volume = {159},
       number = {5},
          eid = {234},
        pages = {234},
          doi = {10.3847/1538-3881/ab880d},
archivePrefix = {arXiv},
       eprint = {2004.03551},
 primaryClass = {astro-ph.EP},
       adsurl = {https://ui.adsabs.harvard.edu/abs/2020AJ....159..234W}
}

@ARTICLE{awiphan2016,
       author = {{Awiphan}, S. and {Kerins}, E. and {Pichadee}, S. and {Komonjinda}, S. and {Dhillon}, V.~S. and {Rujopakarn}, W. and {Poshyachinda}, S. and {Marsh}, T.~R. and {Reichart}, D.~E. and {Ivarsen}, K.~M. and {Haislip}, J.~B.},
        title = "{Transit timing variation and transmission spectroscopy analyses of the hot Neptune GJ3470b}",
      journal = {\mnras},
         year = 2016,
        month = dec,
       volume = {463},
       number = {3},
        pages = {2574-2582},
         doi = {10.1093/mnras/stw2148},
archivePrefix = {arXiv},
       eprint = {1606.02962},
 primaryClass = {astro-ph.EP},
       adsurl = {https://ui.adsabs.harvard.edu/abs/2016MNRAS.463.2574A} 
}

@ARTICLE{edwards2023,
       author = {{Edwards}, Billy and {Changeat}, Quentin and {Tsiaras}, Angelos and {Allan}, Andrew and {Behr}, Patrick and {Hagey}, Simone R. and {Himes}, Michael D. and {Ma}, Sushuang and {Stassun}, Keivan G. and {Thomas}, Luis and {Thompson}, Alexandra and {Boley}, Aaron and {Booth}, Luke and {Bouwman}, Jeroen and {France}, Kevin and {Lowson}, Nataliea and {Meech}, Annabella and {Phillips}, Caprice L. and {Vidotto}, Aline A. and {Yip}, Kai Hou and {Bieger}, Michelle and {Gressier}, Am{\'e}lie and {Janin}, Estelle and {Jiang}, Ing-Guey and {Leonardi}, Pietro and {Sarkar}, Subhajit and {Skaf}, Nour and {Taylor}, Jake and {Yang}, Ming and {Ward-Thompson}, Derek},
        title = "{Characterizing a World Within the Hot-Neptune Desert: Transit Observations of LTT 9779 b with the Hubble Space Telescope/WFC3}",
      journal = {\aj},
         year = 2023,
        month = oct,
       volume = {166},
       number = {4},
          eid = {158},
        pages = {158},
          doi = {10.3847/1538-3881/acea77},
archivePrefix = {arXiv},
       eprint = {2306.13645},
 primaryClass = {astro-ph.EP},
       adsurl = {https://ui.adsabs.harvard.edu/abs/2023AJ....166..158E}
}

@ARTICLE{bai2022,
       author = {{Bai}, Lu and {Gu}, Shenghong and {Wang}, Xiaobin and {Sun}, Leilei and {Kwok}, Chi-Tai and {Hui}, Ho-Keung},
        title = "{The study on transmission spectrum and TTV behaviour of the hot Jupiter WASP-12b}",
      journal = {\mnras},
         year = 2022,
        month = may,
       volume = {512},
       number = {3},
        pages = {3113-3123},
          doi = {10.1093/mnras/stac623},
       adsurl = {https://ui.adsabs.harvard.edu/abs/2022MNRAS.512.3113B}
}

@ARTICLE{agol2005,
       author = {{Agol}, Eric and {Steffen}, Jason and {Sari}, Re'em and {Clarkson}, Will},
        title = "{On detecting terrestrial planets with timing of giant planet transits}",
      journal = {\mnras},
         year = 2005,
        month = may,
       volume = {359},
       number = {2},
        pages = {567-579},
          doi = {10.1111/j.1365-2966.2005.08922.x},
       adsurl = {https://ui.adsabs.harvard.edu/abs/2005MNRAS.359..567A}
}

@ARTICLE{foreman2013,
       author = {{Foreman-Mackey}, Daniel and {Hogg}, David W. and {Lang}, Dustin and {Goodman}, Jonathan},
        title = "{emcee: The MCMC Hammer}",
      journal = {\pasp},
         year = 2013,
        month = mar,
       volume = {125},
       number = {925},
        pages = {306},
          doi = {10.1086/670067},
archivePrefix = {arXiv},
       eprint = {1202.3665},
 primaryClass = {astro-ph.IM},
       adsurl = {https://ui.adsabs.harvard.edu/abs/2013PASP..125..306F}
}

@ARTICLE{gim1995,
       author = {{Gim{\'e}nez}, A. and {Bastero}, M.},
        title = "{A Revision of the Ephemeris-Curve Equations for Eclipsing Binaries with Apsidal Motion}",
      journal = {\apss},
         year = 1995,
        month = apr,
       volume = {226},
       number = {1},
        pages = {99-107},
          doi = {10.1007/BF00626903},
       adsurl = {https://ui.adsabs.harvard.edu/abs/1995Ap&SS.226...99G}
}

@ARTICLE{mannaday2020,
       author = {{Mannaday}, Vineet Kumar and {Thakur}, Parijat and {Jiang}, Ing-Guey and {Sahu}, D.~K. and {Joshi}, Y.~C. and {Pandey}, A.~K. and {Joshi}, Santosh and {Yadav}, Ram Kesh and {Su}, Li-Hsin and {Sariya}, Devesh P. and {Yeh}, Li-Chin and {Griv}, Evgeny and {Mkrtichian}, David and {Shlyapnikov}, Aleksey and {Moskvin}, Vasilii and {Ignatov}, Vladimir and {Va{\v{n}}ko}, M. and {P{\"u}sk{\"u}ll{\"u}}, {\c{C}}.},
        title = "{Probing Transit Timing Variation and Its Possible Origin with 12 New Transits of TrES-3b}",
      journal = {\aj},
         year = 2020,
        month = jul,
       volume = {160},
       number = {1},
          eid = {47},
        pages = {47},
          doi = {10.3847/1538-3881/ab9818},
archivePrefix = {arXiv},
       eprint = {2006.00599},
 primaryClass = {astro-ph.EP},
       adsurl = {https://ui.adsabs.harvard.edu/abs/2020AJ....160...47M}
}

@ARTICLE{patra2017,
       author = {{Patra}, Kishore C. and {Winn}, Joshua N. and {Holman}, Matthew J. and {Yu}, Liang and {Deming}, Drake and {Dai}, Fei},
        title = "{The Apparently Decaying Orbit of WASP-12b}",
      journal = {\aj},
         year = 2017,
        month = jul,
       volume = {154},
       number = {1},
          eid = {4},
        pages = {4},
          doi = {10.3847/1538-3881/aa6d75},
archivePrefix = {arXiv},
       eprint = {1703.06582},
 primaryClass = {astro-ph.EP},
       adsurl = {https://ui.adsabs.harvard.edu/abs/2017AJ....154....4P}
}

@ARTICLE{macie2016,
       author = {{Maciejewski}, G. and {Dimitrov}, D. and {Fern{\'a}ndez}, M. and {Sota}, A. and {Nowak}, G. and {Ohlert}, J. and {Nikolov}, G. and {Bukowiecki}, {\L}. and {Hinse}, T.~C. and {Pall{\'e}}, E. and {Tingley}, B. and {Kjurkchieva}, D. and {Lee}, J.~W. and {Lee}, C.-U.},
        title = "{Departure from the constant-period ephemeris for the transiting exoplanet WASP-12}",
      journal = {\aap},
         year = 2016,
        month = apr,
       volume = {588},
          eid = {L6},
        pages = {L6},
          doi = {10.1051/0004-6361/201628312},
archivePrefix = {arXiv},
       eprint = {1602.09055},
 primaryClass = {astro-ph.EP},
       adsurl = {https://ui.adsabs.harvard.edu/abs/2016A&A...588L...6M}
}

@ARTICLE{macie2021,
       author = {{Maciejewski}, G. and {Fern{\'a}ndez}, M. and {Aceituno}, F. and {Ramos}, J.~L. and {Dimitrov}, D. and {Donchev}, Z. and {Ohlert}, J.},
        title = "{Revisiting TrES-5 b: departure from a linear ephemeris instead of short-period transit timing variation}",
      journal = {\aap},
         year = 2021,
        month = dec,
       volume = {656},
          eid = {A88},
        pages = {A88},
          doi = {10.1051/0004-6361/202142424},
archivePrefix = {arXiv},
       eprint = {2110.14294},
 primaryClass = {astro-ph.EP},
       adsurl = {https://ui.adsabs.harvard.edu/abs/2021A&A...656A..88M}
}

@ARTICLE{yee2020,
       author = {{Yee}, Samuel W. and {Winn}, Joshua N. and {Knutson}, Heather A. and {Patra}, Kishore C. and {Vissapragada}, Shreyas and {Zhang}, Michael M. and {Holman}, Matthew J. and {Shporer}, Avi and {Wright}, Jason T.},
        title = "{The Orbit of WASP-12b Is Decaying}",
      journal = {\apjl},
         year = 2020,
        month = jan,
       volume = {888},
       number = {1},
          eid = {L5},
        pages = {L5},
          doi = {10.3847/2041-8213/ab5c16},
archivePrefix = {arXiv},
       eprint = {1911.09131},
 primaryClass = {astro-ph.EP},
       adsurl = {https://ui.adsabs.harvard.edu/abs/2020ApJ...888L...5Y}
}

@ARTICLE{bouma2020,
       author = {{Bouma}, L.~G. and {Winn}, J.~N. and {Howard}, A.~W. and {Howell}, S.~B. and {Isaacson}, H. and {Knutson}, H. and {Matson}, R.~A.},
        title = "{WASP-4 Is Accelerating toward the Earth}",
      journal = {\apjl},
         year = 2020,
        month = apr,
       volume = {893},
       number = {2},
          eid = {L29},
        pages = {L29},
          doi = {10.3847/2041-8213/ab8563},
archivePrefix = {arXiv},
       eprint = {2004.00637},
 primaryClass = {astro-ph.EP},
       adsurl = {https://ui.adsabs.harvard.edu/abs/2020ApJ...893L..29B}
}

@ARTICLE{mannaday2022,
       author = {{Mannaday}, Vineet Kumar and {Thakur}, Parijat and {Southworth}, John and {Jiang}, Ing-Guey and {Sahu}, D.~K. and {Mancini}, L. and {Va{\v{n}}ko}, M. and {Kundra}, Emil and {Gajdo{\v{s}}}, Pavol and {A-thano}, Napaporn and {Sariya}, Devesh P. and {Yeh}, Li-Chin and {Griv}, Evgeny and {Mkrtichian}, David and {Shlyapnikov}, Aleksey},
        title = "{Revisiting the Transit Timing Variations in the TrES-3 and Qatar-1 Systems with TESS Data}",
      journal = {\aj},
         year = 2022,
        month = nov,
       volume = {164},
       number = {5},
          eid = {198},
        pages = {198},
          doi = {10.3847/1538-3881/ac91c2},
archivePrefix = {arXiv},
       eprint = {2209.04080},
 primaryClass = {astro-ph.EP},
       adsurl = {https://ui.adsabs.harvard.edu/abs/2022AJ....164..198M}
}

@ARTICLE{fulton2018,
       author = {{Fulton}, Benjamin J. and {Petigura}, Erik A. and {Blunt}, Sarah and {Sinukoff}, Evan},
        title = "{RadVel: The Radial Velocity Modeling Toolkit}",
      journal = {\pasp},
         year = 2018,
        month = apr,
       volume = {130},
       number = {986},
        pages = {044504},
          doi = {10.1088/1538-3873/aaaaa8},
archivePrefix = {arXiv},
       eprint = {1801.01947},
 primaryClass = {astro-ph.IM},
       adsurl = {https://ui.adsabs.harvard.edu/abs/2018PASP..130d4504F}
}

@misc{10.17909/t9-nmc8-f686,
doi = {10.17909/T9-NMC8-F686},
url = {http://archive.stsci.edu/doi/resolve/resolve.html?doi=10.17909/t9-nmc8-f686},
author = {{TESS Team}},
title = {TESS Light Curves - All Sectors},
publisher = {STScI/MAST},
year = {2021}
}
\bibliographystyle{aasjournal}



\end{document}